\documentclass[draft]{agujournal2019}
\usepackage{url} 
\usepackage{soul}
\usepackage[utf8]{inputenc}  
\usepackage{bm}

\draftfalse

\journalname{Journal of Advances in Modeling Earth Systems (JAMES)}

\begin{document}

%
%


\title{Efficient optimization of a regional water elevation model with an automatically generated adjoint}

%
%




\authors{Tuomas K\"arn\"a\affil{1}\thanks{Current affiliation: Intel Corporation}, Joseph G. Wallwork\affil{2}, Stephan C. Kramer\affil{2}}

\affiliation{1}{Finnish Meteorological Institute, Helsinki, Finland}
\affiliation{2}{Imperial College, London, United Kingdom}




\correspondingauthor{Tuomas K\"arn\"a}{tuomas.karna@protonmail.com}




\begin{keypoints}
\item Adjoint-based optimization is used to optimize bottom friction coefficient in 2D water elevation model for the Baltic Sea
\item The discrete adjoint model is automatically generated by leveraging a symbolic representation of the discrete forward model equations
\item The optimization method is robust and results in significant improvement in the sea surface height performance at tide gauge locations
\end{keypoints}

%
%

%
%


\begin{abstract}
Calibration of unknown model parameters is a common task in many ocean model applications. We present an adjoint-based optimization of an unstructured mesh shallow water model for the Baltic Sea. Spatially varying bottom friction parameter is tuned to minimize the misfit with respect to tide gauge sea surface height (SSH) observations. A key benefit of adjoint-based optimization is that computational cost does not depend on the number of unknown variables. Adjoint models are, however, typically very laborious to implement. In this work, we leverage a domain specific language framework in which the discrete adjoint model can be obtained automatically. The adjoint model is both exactly compatible with the discrete forward model and computationally efficient. A gradient-based quasi-Newton method is used to minimize the misfit. Optimizing spatially-variable parameters is typically an under-determined problem and can lead to over-fitting. We employ Hessian-based regularization to penalize the spatial curvature of the friction field to overcome this problem. The SSH dynamics in the Baltic Sea are simulated for a 3-month period. Optimization of the bottom friction parameter results in significant improvement of the model performance. The results are especially encouraging in the complex Danish Straits region, highlighting the benefit of unstructured meshes. Domain specific language frameworks enable automated model analysis and provide easy access to adjoint modeling. Our application shows that this capability can be enabled with few efforts, and the optimization procedure is robust and computationally efficient.
\end{abstract}

\section*{Plain Language Summary}


Ocean circulation models have several unknown parameters that must be tuned for each application in order to produce physically meaningful results. The tuning process can be a very laborious and time consuming task. In this paper, we investigate an automated way to tune the model's friction at the sea bed to minimize the model's error in predicted sea surface height. The method is based on a novel way of defining the model's equations which enables solving such optimization problems automatically. The methodology is tested in the Baltic Sea. The modeled sea surface height is compared against observations at several tide gauges. We show that by optimizing the bottom friction, the model's capability to predict sea surface height improves significantly. Moreover, we show that the optimization process is robust and computationally efficient.

%
%

\section{Introduction}

Numerical modeling is an indispensable tool in oceanography and climate sciences.
One of its major bottlenecks, however, is fitting the model state to observations.
Model configuration must be carefully tuned and calibrated to replicate observational data.
In addition to merely running simulations, one typically also needs to quantify the uncertainty of model predictions \cite{kalmikov2014,loose2021}, or estimate unknown parameters \cite{heemink2002,zaron2011,zhang2011,almeida2018,warder2022}.
The bottom friction coefficient, for example, is commonly regarded as an unknown free parameter in coastal applications (e.g.,~\citeA{zhang2011}).
In addition, as the amount of high-resolution observational data increases, sophisticated data assimilation methods are needed to synthesize the observation data and fill in the gaps with a physically meaningful ocean state.
All of these use cases are examples of inverse modeling.

Adjoint models provide an efficient way to solve inverse problems.
In short, adjoint models allow evaluating the gradient of model outputs with respect to the model's internal state, forcing fields, or parameters.
The drawback is that implementing the adjoint model is technically challenging and labor intensive \cite{marotzke1999,heemink2002,vidard2015}.
The adjoints of the ROMS and NEMO ocean models, for example, have been implemented manually, by differentiating each operator of the model \cite{moore2004,vidard2015}.
Maintaining such a hard-coded adjoint, however, requires constant human intervention as the models evolve; both ROMS and NEMO adjoint models now lag behind the latest forward model versions.
In the case of the MITgcm model, the adjoint has been derived via Automatic Differentiation (AD), i.e.~by automatically differentiating the Fortran source code \cite{marotzke1999}.
The MITgcm adjoint has been used in a wide variety of applications, but despite its success, the AD-generated adjoint approach has not been widely adopted in other models.
Therefore, for the majority of ocean models, adjoint capability is not available and inverse problems are difficult to solve, e.g.~one has to rely on ensemble runs to obtain statistical estimates of the model's sensitivity.
In this paper, we demonstrate that by leveraging suitable high-level abstractions and code generation, the adjoint model can be generated and solved automatically, with very few efforts once the forward model has been implemented.

Adjoint models come in two flavors, continuous and discrete adjoints.
In the former, the adjoint equations are derived directly from the continuous model equations and then discretized with a method of choice (e.g.~finite volume (FV) or finite element (FE) method).
In the discrete adjoint case, on the other hand, one differentiates the discrete (e.g.~FV or FE) equations to derive an adjoint that is exactly compatible with the discrete forward model.
Differentiating the model source code, as in the case of AD, also results in a discrete adjoint.
There are advantages and disadvantages to both continuous and discrete adjoint approaches \cite{sirkes1997}.
For example, the continuous adjoint approach is flexible in the sense that the user is able to choose different discretization methods for the forward and adjoint equations, but comes with the burden of being tedious and error-prone to implement.
The discrete adjoint approach, on the other hand, is less flexible, but requires less user effort and produces gradients that are exactly compatible with the discretized forward model.
This is greatly beneficial in applications -- such as the one considered in this paper -- where gradient-based optimization methods are used, since the convergence of such methods is ensured.

In this work, we use an unstructured mesh FE ocean model, Thetis \cite{karna2018}, for which the discrete adjoint model can be derived automatically.
Thetis has been implemented in the generic Firedrake FE modeling framework \cite{rathgeber2016}.
Firedrake uses a domain specific language (Unified Form Language, \cite{alnaes2014}) to describe the FE weak forms.
An automated code generator \cite{homolya2018} is then used to generate C code to evaluate the terms of the weak form.
The equations are assembled as a linear system and solved in parallel with the PETSc solver library \cite{petsc2021}.
Leveraging the symbolic representation, it is possible to derive the discrete adjoint model automatically by differentiating the symbolic FE equations \cite{farrell2013}.
The pyadjoint library handles the taping of the operator calls, solving the adjoint equation and evaluating the gradient of the cost function.
Finally, a gradient-based quasi-Newton method is used to solve the optimization problem.

We use a discontinuous Galerkin FE 2D shallow water model, implemented within the Thetis framework, to simulate tidal and atmospherically-driven water elevation dynamics in the Baltic Sea.
Our primary focus is on the Baltic Sea and the Danish Straits, but as the North Sea is tightly coupled to the dynamics, the two seas must be simulated as a single dynamical system \cite{daewel2013,hordoir2019,karna2021}.
We use the adjoint model to optimize the model's bottom friction coefficient to improve the representation of water elevation dynamics with respect to tide gauge observations.

The two seas exhibit quite different water elevation dynamics.
The North Sea is a tidal system with dominant semi-diurnal tides \cite{huthnance1991}.
Tidal range varies from over 6 m in the English channel to roughly 0.4 m in Skagerrak.
Tides are effectively filtered out in the Danish waters and only weak tides ($<$10 cm range) are observed in the Baltic Sea.
The atmospherically-induced water elevation gradient across the Danish Straits controls the water volume exchange to and from the Baltic Sea in synoptic and longer time scales \cite{mohrholz2015,grawe2015,mohrholz2018}.
Consequently, water elevation in the Baltic Sea is mainly governed by episodic filling and emptying of the basin, storm surges, as well as atmospherically-generated seiche oscillations whose typical periods range from 17 to 31 h \cite{lepparanta2009}.

Previously, adjoint models have been used in several coastal ocean applications.
\citeA{massmann2010} used an AD-based adjoint of an unstructured mesh finite element model to study the sensitivity of a North Sea water elevation model versus the bottom friction coefficient, bathymetry, and open boundary forcing.
They found that the model skill was most sensitive to bottom friction, although in certain regions there was some overlap with sensitivity to bathymetry.
Similarly, \citeA{warder2021} studied the spatial and temporal sensitivies of a storm surge model of the North Sea with respect to bottom friction, bathymetry and wind forcing using Thetis.

\citeA{almeida2018} use the Delft3D-FLOW model to simulate a section of the Columbia River. A continous adjoint was implemented for the model and used to optimize the model's bathymetry based on surface velocity measurements. As the continuous adjoint model does not match exactly with the discrete forward model, a careful solution strategy was devised to avoid numerical instabilities.

\citeA{heemink2002} presented an adjoint-based inverse model of a 3D hydrostatic model and used it to optimize bottom friction, vertical viscosity coefficient and bathymetry in an application to the European Continental Shelf. The adjoint model was implemented manually. A common problem in parameter estimation is that the problem is typically underdetermined: in the case of a spatially-varying parameter field, there are far too many degrees of freedom compared to the amount of observation data. \citeA{heemink2002} reduced the dimensionality of the problem by allowing the parameter fields to vary only in predefined subdomains of the model grid.
The choice of subdomains was informed by inspecting the gradient of the cost function, calculated with the adjoint, and also expert knowledge of the system dynamics (e.g.~amphidromic points).

\citeA{zhang2011} used a shallow water adjoint model to optimize constant and
spatially-variable bottom friction coefficient for the Bohai Sea and Yellow Sea.
Also here, the dimensionality of the problem was reduced by optimizing the friction values only in a sparse subset of the model grid
points and using linear interpolation in between.

The availability of adjoint-based gradient information has a large number of other potentially powerful applications.
In coastal engineering applications the adjoint can be used to optimize the design of engineering structures and their interaction with the environment.
As an example, the aforementioned pyadjoint approach has been used to study the optimal placement of a large number of tidal turbines in \cite{funke2014, funke2016, goss2021}.
The automated adjoint approach can easily be extended to coupled model applications -- as demonstrated in \cite{clare2022} where inversion and sensitivity analyses where performed with Thetis coupled to a morphodynamic model.
Finally, \citeA{wallwork2020} demonstrated that adjoint-based goal-oriented error estimate can be used for mesh adaptation to optimize the model accuracy for a specified quantity of interest (e.g.~power output of tidal turbines).

Compared to previous adjoint-based friction optimization applications, the novelty of the present work is the use of the automatically derived adjoint.
This approach offers significant benefits to the user, as one only needs to implement the forward model and the cost function.
This capability is built-in in Firedrake and therefore immediately available to any Firedrake-based implementation.
In addition, we have implemented additional functionality in Thetis to provide the observation-based cost function, adjoint model related file IO -- and the final missing ingredient -- coupling with a quasi-Newton optimization method.

We use regularization to constrain the under-determined bottom friction optimization problem. That is, an additional term is added in the cost function to penalize the second derivatives of the friction field to control its smoothness.
We show that the optimization procedure works well in a large and complex coastal domain where SSH dynamics vary across several time scales and the adjoint solver is computationally efficient.

The 2D shallow water model and its discretization is presented in Section \ref{sec:shallow_water_model}.
The variational inverse problem and its application to bottom friction optimization is presented in Section \ref{sec:variational_inverse_problem}.
Section \ref{sec:baltic_configuration} outlines the model configuration for the North Sea-Baltic Sea, followed by results in Section \ref{sec:results}. Discussion and conclusions are presented in Sections \ref{sec:discussion} and \ref{sec:conclusions}, respectively.
Further details on the forward and inverse modelling techniques are provided in \ref{sec:p2_bathy} and \ref{sec:hessian_recovery}.

\section{Shallow-water model} \label{sec:shallow_water_model}

Denoting the free surface elevation and depth averaged velocity by $\eta$ and $\textbf{u} = (u, v)$, respectively,
the shallow water equations in non-conservative form read
\begin{eqnarray}
 \frac{\partial \eta}{\partial t} + \nabla \cdot (H \textbf{u}) &=& 0 \label{eq:free_surface} \\
 \frac{\partial \textbf{u}}{\partial t} +
 \textbf{u} \cdot \nabla\textbf{u} +
 f\textbf{e}_z \wedge \textbf{u} +
 g \nabla \eta + \frac{1}{\rho_0} \nabla p_a
 &=& \textbf{D}_{\textbf{u}}
 + \frac{\bm{\tau}_w + \bm{\tau}_b}{H \rho_0}, \label{eq:momentum}
\end{eqnarray}
where
$H = \eta + h$ is the total water column depth, $h$ is the bathymetry,
$f$ is the Coriolis parameter, $\textbf{e}_z$ an upward vertical unit vector,
$g$ the gravitational acceleration, $\rho_0$ the constant reference water density,
$p_a$ the atmospheric pressure,
$\textbf{D}_{\textbf{u}} = \nabla \cdot \big( \nu ( \nabla \textbf{u} + (\nabla \textbf{u})^T )\big)$ is the viscosity operator and $\nu$ is the horizontal eddy viscosity,
$\bm{\tau}_w$ and $\bm{\tau}_b$ denote the surface (wind) and bottom stresses, respectively.

In this work, we use the wind stress formula by \citeA{large2008}:
\begin{eqnarray}
 \bm{\tau}_w &=& C_D^w |\textbf{u}_w| \textbf{u}_w, \\
 C_D^w &=& \left\{
     \begin{array}{rl}
     a_1 \frac{1}{|\textbf{u}_w|} + a_2 + a_3 |\textbf{u}_w| + a_8 |\textbf{u}_w|^6, & |\textbf{u}_w| < 33\ \mathrm{m/s}, \\
     0.00234, & \mathrm{otherwise}
     \end{array}
    \right.
\end{eqnarray}
where $\textbf{u}_w$ stands for the wind velocity at 10 m height; the $a_i$
coefficients are defined in \citeA{large2008}.

The bottom friction is parametrized by the Manning formula,
\begin{eqnarray}
 \bm{\tau}_b &=& C_D |\textbf{u}| \textbf{u}, \label{eq:bottom_friction} \\
 C_D &=& g \frac{\mu^2}{H^{1/3}},
\end{eqnarray}
where $\mu$ is the Manning friction coefficient.
Generally $\mu$ can be regarded as an unknown, spatially-variable, model and mesh-dependent parameter.

\subsection{Finite element discretization}

We use the Thetis model implementation of the shallow water equations \cite{karna2018}.
The 2D model domain and its boundary are denoted by $\Omega$ and $\Gamma$, respectively.
The boundary consists of open ocean ($\Gamma_o$) and closed land ($\Gamma_c$) parts, i.e.~$\Gamma = \Gamma_o \cup \Gamma_c$.

\begin{figure}
 \centering
 \noindent\includegraphics[width=0.6\textwidth]{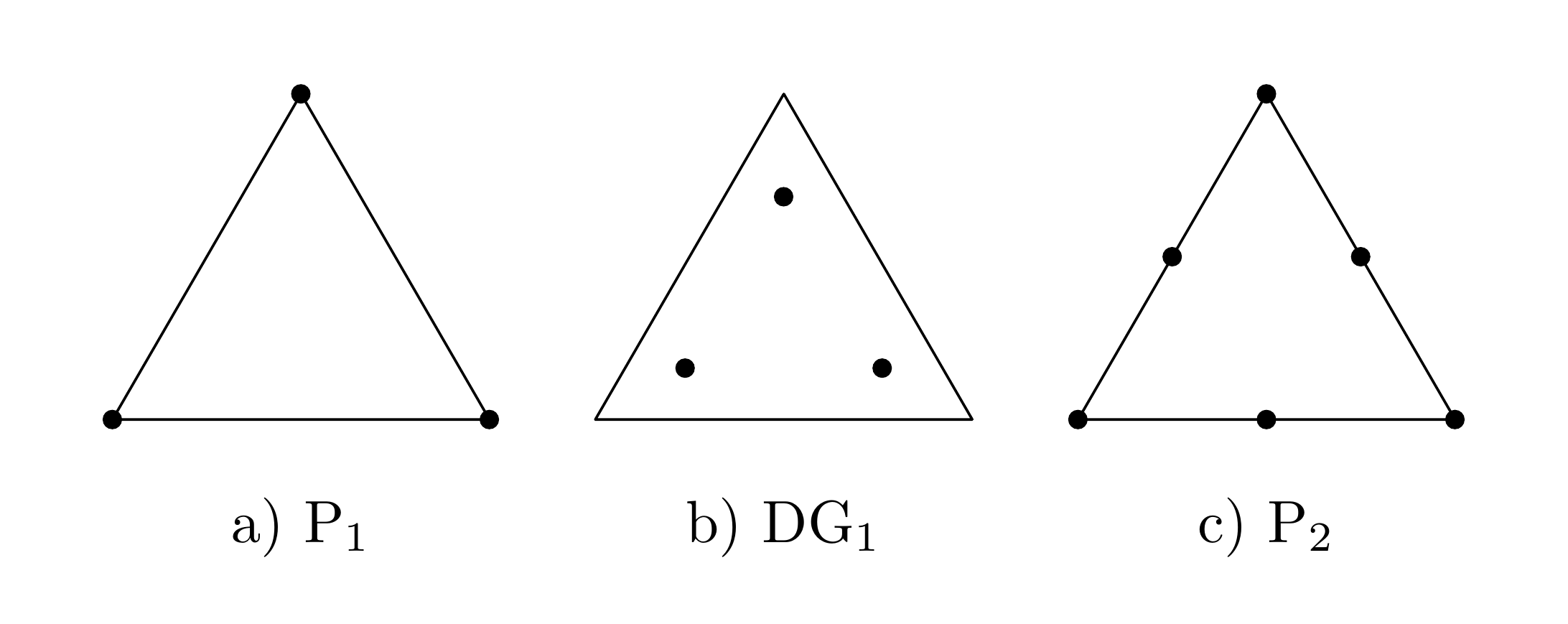}
 \caption{
  Finite elements:
  (a) linear continuous;
  (b) linear discontinuous; and
  (c) quadratic continuous scalar element.
  The dots denote scalar degrees of freedom. For vector-valued fields, each component is stored as a scalar.
 }
 \label{fig:elements}
\end{figure}

Thetis supports several finite element families. In this work,
the equations are discretized with linear discontinuous (DG$_1$) finite elements (see Fig. \ref{fig:elements}).
The same element is used for both of the prognostic fields, $\eta$ and $\textbf{u}$.
Let $\phi$ and $\bm{\phi}$ denote the scalar and vector valued test functions in the DG$_1$ function space.
The weak form of the shallow water problem is obtained by multiplying the governing equations (\ref{eq:free_surface}-\ref{eq:momentum}) by the test functions $\phi$ and $\bm{\phi}$, respectively and
integrating over the domain $\Omega$:

\begin{eqnarray}
 \int_\Omega \frac{\partial \eta}{\partial t} \phi \;\mathrm dx + \mathcal{S}_{div} &=& 0, \quad \forall \phi \label{eq:weak_freesurf}\\
 \int_\Omega \frac{\partial \mathbf{u}}{\partial t} \cdot \bm{\phi} \;\mathrm dx
 + \mathcal{S}_{adv}
 + \mathcal{S}_{cor}
 + \mathcal{S}_{pg}
 + \mathcal{S}_{pa}
 &=&
 \mathcal{S}_{visc}
 + \mathcal{S}_{\bm{\tau}}, \quad \forall \bm{\phi}. \label{eq:weak_momentum}
\end{eqnarray}
\noindent The bilinear forms of each term, $\mathcal{S}_{\cdot}$, are defined below.

Let $\mathcal{T}$ stand for the triangulation of the domain $\Omega$.
The set of element interfaces is denoted by $\mathcal{I} = \{k \cap k' \vert k,k' \in \mathcal{T}\}$, and $\mathbf{n}=(n_x,n_y)$ denotes the outward unit normal vector of an interface $e\in\mathcal{I}$.
On the interfaces, the DG$_1$ functions are discontinuous and do not have a unique value.
We define the average and jump operators,
\begin{eqnarray}
 \{\!\!\{a\}\!\!\} &=& \frac{1}{2}(a^+ + a^-), \\
 {[}\!{[} a \textbf{n} {]}\!{]} &=& a^+ \mathbf{n}^+ + a^- \mathbf{n}^-, \\
 {[}\!{[}\mathbf{u} \cdot \mathbf{n} {]}\!{]} &=& \mathbf{u}^+ \cdot \mathbf{n}^+ + \mathbf{u}^- \cdot \mathbf{n}^-, \\
 {[}\!{[} \mathbf{u} \mathbf{n} {]}\!{]} &=& \mathbf{u}^+ \mathbf{n}^+ + \mathbf{u}^- \mathbf{n}^-,
\end{eqnarray}
where the superscripts `$+$' and `$-$' arbitrarily label the values on
either side of the interface, and $\mathbf{n}^- = -\mathbf{n}^+$.

The $H\mathbf{u}$ divergence term is integrated by parts:

\begin{eqnarray}
 \mathcal{S}_{div} = -\int_\Omega H (\textbf{u}\cdot\nabla \phi) \;\mathrm dx
  + \int_{\mathcal{I}} (H^* \textbf{u}^*) \cdot {[}\!{[} \phi \textbf{n} {]}\!{]} \;\mathrm dS
  + \int_{\Gamma_o} (H \textbf{u}) \cdot \phi \textbf{n} \;\mathrm dS
\end{eqnarray}
\noindent where $H^*$ and $\textbf{u}^*$ denote the interface terms obtained from an approximate Riemann solver defined below.
Note that the $H\mathbf{u}$ flux vanishes on the closed boundary $\Gamma_c$.

The advection term is also integrated by parts (also here the flux across $\Gamma_c$ vanishes):

\begin{eqnarray}
 \mathcal{S}_{adv}
        = - \int_\Omega \nabla_h \cdot (\textbf{u} \bm{\phi}) \cdot \textbf{u} \;\mathrm dx
        + \int_{\mathcal{I}} \{\!\!\{ \textbf{u} \}\!\!\} \cdot {[}\!{[} \bm{\phi}
        (\textbf{u}\cdot\textbf{n}) {]}\!{]} \;\mathrm dS
        + \int_{\Gamma_o} \textbf{u} \cdot \bm{\phi}
        (\textbf{u}\cdot\textbf{n}) \;\mathrm dS.
\end{eqnarray}

The Coriolis and atmospheric pressure gradient terms read

\begin{eqnarray}
 \mathcal{S}_{cor} &=& \int_\Omega f\textbf{e}_z \wedge \textbf{u} \cdot \bm{\phi} \;\mathrm dx, \\
 \mathcal{S}_{pa} &=& \int_\Omega \frac{1}{\rho_0} \nabla p_a \cdot \bm{\phi} \;\mathrm dx.
\end{eqnarray}

The pressure gradient term is integrated by parts

\begin{eqnarray}
 \mathcal{S}_{pg}
        = - \int_\Omega g \eta \nabla \cdot \bm{\phi} \;\mathrm dx
        + \int_{\mathcal{I}} g \eta^* {[}\!{[} \bm{\phi} \cdot \textbf{n} {]}\!{]} \;\mathrm dS
        + \int_\Gamma g \eta \bm{\phi} \cdot \textbf{n} \;\mathrm dS.
\end{eqnarray}

The viscosity operator is discretized with the interior penalty method \cite{riviere2008,hillewaert2013}:

\begin{eqnarray}
 \mathcal{S}_{visc} &=& \int_\Omega (\nabla \boldsymbol{\psi}) : \boldsymbol{\tau}_\nu \;\mathrm dx
        - \int_{\mathcal{I}} {[}\!{[} \boldsymbol{\psi} \textbf{n} {]}\!{]} \cdot \{\!\!\{ \boldsymbol{\tau}_\nu \}\!\!\} \;\mathrm dS \nonumber \\
        && - \int_{\mathcal{I}} \{\!\!\{ \nu_h \}\!\!\}
        {[}\!{[} \mathbf{u} \textbf{n} {]}\!{]} \cdot \{\!\!\{ \nabla \boldsymbol{\psi} \}\!\!\} \;\mathrm dS
        + \int_{\mathcal{I}} \sigma \{\!\!\{ \nu_h \}\!\!\} {[}\!{[} \mathbf{u} \textbf{n} {]}\!{]} \cdot
        {[}\!{[} \boldsymbol{\psi} \textbf{n} {]}\!{]} \;\mathrm dS \nonumber \\
        &&
        - \int_\Omega \frac{\nu_h \nabla(H)}{H} \cdot ( \nabla \mathbf{u} + (\nabla \mathbf{u})^T ) \cdot \bm{\phi} \;\mathrm dx
\end{eqnarray}
\noindent where $\boldsymbol{\tau}_\nu = \nu_h \nabla \mathbf{u}$ denotes the viscous flux.
The colon operator, $:$, stands for the  double inner product that contracts over both indices of the tensor expressions.
The last term is an additional source term that accounts for the bathymetry gradient.
The viscous flux is zero on $\Gamma_c$ and on $\Gamma_o$.

Finally, the surface and bottom stress terms are given by

\begin{eqnarray}
 \mathcal{S}_{\bm{\tau}} = \int_\Omega \frac{\bm{\tau}_w + \bm{\tau}_b}{H \rho_0} \cdot \bm{\phi} \;\mathrm dx.
\end{eqnarray}

In this work, we use the Roe fluxes \cite{roe1981,leveque2002} to stabilize the $\mathcal{S}_{div}$ and $\mathcal{S}_{pg}$ terms:

\begin{eqnarray}
 \eta^* &=& \{\!\!\{ \eta \}\!\!\} + \sqrt{\frac{g}{H}} {[}\!{[} \mathbf{u} \cdot \mathbf{n} {]}\!{]} \\
 \mathbf{u}^* &=& \{\!\!\{ \mathbf{u} \}\!\!\} + \sqrt{\frac{H}{g}} {[}\!{[} \eta \mathbf{n} {]}\!{]} \\
 H^* &=& \eta^* + h.
\end{eqnarray}

The spatial integrals are evaluated with a standard Gauss quadrature rule of degree 3.
Denoting the individual DG$_1$ basis functions by $\phi_i$ and $\bm{\phi}_i$, the discrete FE representation of
$\eta$ and $\textbf{u}$ fields are $\eta^h := \sum_i \eta_i \phi_i$ and
$\textbf{u}^h := \sum_i \mathbf{u}_i \bm{\phi}_i$, where $\eta_i$ and $\mathbf{u}_i$ are the nodal values.
The superscript $h$ is used to denote the discrete FE fields in contrast to the continuous ones in (\ref{eq:free_surface}-\ref{eq:momentum}).
The model state (dual) vectors are $\bm{\eta}^h = (\eta_1, \eta_2, \ldots)$,
$\mathbf{u}^h = (u_1, u_2, \ldots, v_1, v_2, \ldots)$; the concatenated state vector is denoted by $\mathbf{q}^h = (\mathbf{u}^h, \mathbf{\eta}^h)$.
Replacing the test functions in (\ref{eq:weak_freesurf}-\ref{eq:weak_momentum}) by $\phi_i$ and $\bm{\phi}_i$, respectively, the equations can be written in a vector form:

\begin{eqnarray}
 \underline{\mathbf{M}} \frac{\partial \mathbf{q}^h}{\partial t} &=& \mathbf{S}(\mathbf{q}^h),
\end{eqnarray}
\noindent where $\underline{\mathbf{M}}$ denotes the 3-block DG$_1$ mass matrix, i.e.~$\underline{\mathbf{M}} = \mathrm{diag}(\underline{\mathbf{M}}_{\phi}, \underline{\mathbf{M}}_{\phi}, \underline{\mathbf{M}}_{\phi})$ and $[\underline{\mathbf{M}}_{\phi}]_{i.j} = \int_\Omega \phi_i \phi_j dx$, and $\mathbf{S}$ denotes the remaining bilinear forms. Hereafter we omit the $h$ superscript for brevity.

\subsection{Time integration}

The solution is marched in time with a fully implicit Runge-Kutta
scheme. In this work, we use the two-stage, 2nd order accurate Diagonally
Implicit Runge Kutta method DIRK(2,2), defined by the Butcher tableau
\cite{ascher1997},

\begin{eqnarray}
\begin{array}{c|cc}
 c_1 & a_{1,1} & 0 \\
 c_2 & a_{2,1} & a_{2,2} \\
\hline
 & b_1 & b_2
 \end{array}
 & &
\begin{array}{c|cc}
 \gamma & \gamma & 0 \\
 1 & 1-\gamma & \gamma \\
\hline
  & 1-\gamma & \gamma
 \end{array}
\end{eqnarray}
\noindent with $\gamma = (2-\sqrt{2})/2$.

Denoting the time step by $\Delta t$, solution at time $t^n$ by $\textbf{q}^n$,
and intermediate solutions by $\mathbf{q}^{(i)}$, the $m$-stage DIRK
iteration reads

\begin{eqnarray}
 \mathbf{q}^{(0)} &=& \mathbf{q}^n, \\
 \underline{\mathbf{M}} \mathbf{q}^{(i)} &=& \underline{\mathbf{M}} \mathbf{q}^{(i-1)} + \Delta t \sum_{j = 1}^{i} a_{i,j} \mathbf{S}(\mathbf{q}^{(j)}), \quad \forall i=1,\ldots,m, \\
 \mathbf{q}^{n+1} &=& \mathbf{q}^{(m)},
\end{eqnarray}

For brevity, we denote the entire nonlinear forward model update operator by $F$, i.e.

\begin{eqnarray}
 F(\mathbf{q}^{n+1}, \mathbf{q}^{n}) &=& 0. \label{eq:fwd_update}
\end{eqnarray}

Agglomerating the model state over all $N$ time steps into a single vector
$\mathbf{Q} = (\mathbf{q}^0,\ldots,\mathbf{q}^N)$, we can re-write the entire
forward operator as

\begin{eqnarray}
 \mathcal{F}(\mathbf{Q}) = 0. \label{eq:forward}
\end{eqnarray}

Note that the $\mathcal{F}$ operator is only introduced to simplify notation, it is never assembled.
Indeed, $\mathcal{F}$ consists of nonlinear solves and it can only be evaluated by iterating over the time steps.
As seen in (\ref{eq:fwd_update}), the forward update depends on past values of the model state.
Let $\tilde{\mathcal{F}}$ denote a linearized forward operator. Due to the time dependency,
$\tilde{\mathcal{F}}$ could be assembled into a lower-diagonal matrix operator.
In what follows, we'll see that the adjoint model requires a transpose of $\tilde{\mathcal{F}}$, therefore reversing the time dependency.

\section{Variational inverse problem} \label{sec:variational_inverse_problem}

\subsection{General description} \label{subsec:general_inv_problem}

\begin{figure}
 \centering
 \noindent\includegraphics[width=0.8\textwidth]{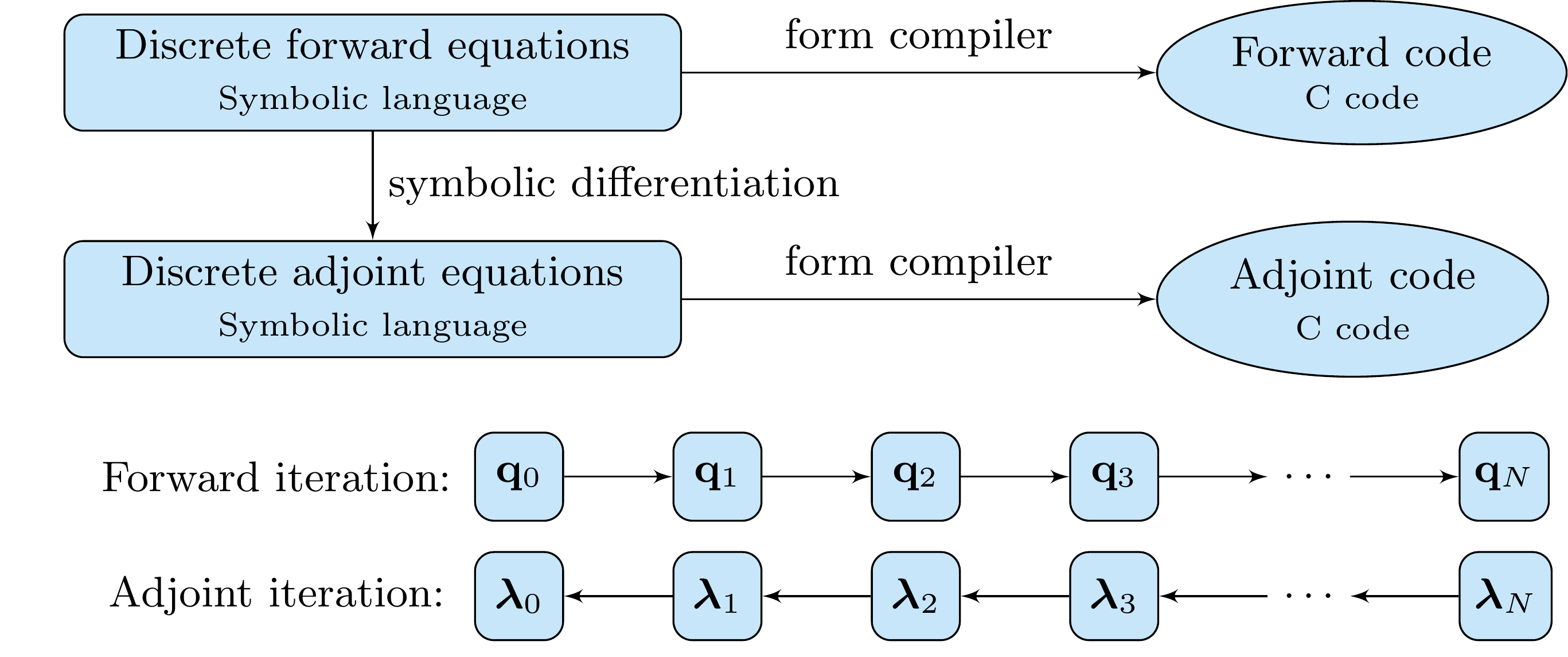}
 \caption{
  Automated generation of the discrete adjoint model.
  The forward model updates the model state, $\mathbf{q}_i$, forward in time while the
  adjoint model iterates the adjoint variable, $\boldsymbol{\lambda}_i$, backward in time.
 }
 \label{fig:adjoint}
\end{figure}

The forward model depends on some unknown control parameters, $\bm{\theta}$, that need to be estimated, i.e.~$\mathcal{F} = \mathcal{F}(\mathbf{Q}, \bm{\theta})$.
Let $J$ denote the user-defined, scalar-valued cost function we aim to minimize.
The inverse optimization problem then reads
\begin{eqnarray}
 \min_{\mathbf{Q},\bm{\theta}}\, J(\mathbf{Q},\bm{\theta}), \quad
 \textrm{subject to}\,\, \mathcal{F}(\mathbf{Q}, \bm{\theta}) = 0. \label{eq:optimization_problem}
\end{eqnarray}

The minimization problem can be solved with standard gradient-based optimization methods provided that one can compute the gradient of $J$. Following the notation of \cite{funke2013} (where $\mathbf{Q}$ and $\bm{\theta}$ are defined as column vectors but the partial derivatives of a scalar, e.g., ${\partial J}/{\partial{\mathbf Q}}$, are row vectors), the gradient can be written as:

\begin{eqnarray}
 \frac{\mathrm dJ}{\mathrm d\bm{\theta}} = \frac{\partial J}{\partial \mathbf{Q}} \frac{\mathrm d\mathbf{Q}}{\mathrm d\bm{\theta}} + \frac{\partial J}{\partial \bm{\theta}}.
\end{eqnarray}

The latter term on the right hand side can be obtained by differentiating $J$ symbolically.
The first term is unknown and can be computed from the discrete adjoint equation,

\begin{eqnarray}
 \left[\frac{\partial \tilde{\mathcal{F}}}{\partial \mathbf{Q}}\right]^T\mathbf{L} = \left[\frac{\partial J}{\partial \mathbf{Q}}\right]^T, \label{eq:adjoint}
\end{eqnarray}
\noindent where $\mathbf{L} = (\boldsymbol{\lambda}^0, \boldsymbol{\lambda}^1, \ldots, \boldsymbol{\lambda}^N)$ is the time-agglomerated adjoint variable.
The adjoint variables, $\boldsymbol{\lambda}^n$, have the same dimension as the model state, $\textbf{q}^n$.
That is, for the model state variables, $\eta^n$ and $\textbf{u}^n$, we have corresponding spatially and temporally varying adjoint fields.
Note that due to the transpose, the adjoint equation must be solved backward in time (Fig. \ref{fig:adjoint}).

Once $\mathbf{L}$ is known, the gradient of $J$ can be computed:

\begin{eqnarray}
 \frac{\mathrm dJ}{\mathrm d\bm{\theta}} &=&
 -\mathbf{L}^T \frac{\partial \tilde{\mathcal{F}}}{\partial \bm{\theta}} + \frac{\partial J}{\partial \bm{\theta}}.
\end{eqnarray}

The adjoint method is attractive as the cost of solving the adjoint equation, and hence its evaluation of the gradient, does not depend on the number of unknown parameters, $\bm{\theta}$ \cite{funke2013}.
This is in contrast to many other methods of calculating gradient information such as finite difference approximations, the tangent linear model, or ensemble methods, where computational cost increases with the number of control variables if evaluation of the full gradient is required.

With the pyadjoint library, the adjoint equation can be derived automatically \cite{farrell2013}.
Leveraging symbolic representation of the finite element discretization, (\ref{eq:forward}),
the adjoint equation is formed by differentiating the forward equations (Fig.\ref{fig:adjoint}); a similar procedure is used to compute ${\partial \tilde{\mathcal{F}}}/{\partial \bm{\theta}}$ and ${\partial J}/{\partial \bm{\theta}}$.

The time-dependent forward problem is solved first and pyadjoint is used to record the sequence of forward solve operations on ``tape'' (also known as the Wengert list, see e.g., \citeA{griewank2008}) over the range of $N$ time steps.
The adjoint equation (\ref{eq:adjoint}) is solved backward in time by rewinding the tape and applying the linearized adjoint operators (Fig.\ref{fig:adjoint}).
In this work, the forward model state vectors, $\boldsymbol{Q}$, are kept in memory and retrieved during the backward pass.
For larger problems, this may exceed the amount of physical memory and $\boldsymbol{Q}$ must be stored to disk. Efficient pyadjoint checkpointing mechanism is under development.
Once the adjoint variables, $\boldsymbol{\lambda}^n$, are known, ${\mathrm dJ}/{\mathrm d\bm{\theta}}$ can be evaluated.

Automated generation of the discrete adjoint model circumvents two major bottlenecks in adjoint modeling:
First, implementing the adjoint model by hand is tedious, often comparable to the cost of implementing the forward model itself.
Second, we utilize the same code generator as with the forward model to obtain low-level implementation for the adjoint model (Fig. \ref{fig:adjoint}).
Thus the adjoint is computationally efficient, as the code generator can apply similar code optimization methods, such as loop unrolling or tiling, to the adjoint kernels.

Firedrake and the pyadjoint library provide an interface to derive and solve the adjoint model automatically. The user only needs to define the forward model $\mathcal{F}$ and the cost function $J$. The gradient ${\partial J}/{\partial \bm{\theta}}$ can then be evaluated efficiently with only few modifications to the model code, enabling a wide range of inverse modeling applications.

Utilizing the gradient ${\mathrm dJ}/{\mathrm d\bm{\theta}}$, the optimization problem (\ref{eq:optimization_problem}) is solved with a quasi-Newton method where the Hessian of $J$ is estimated during the iteration.
The convergence of the quasi-Newton method depends on several factors, e.g. nonlinearity of the problem and smoothness of the cost function.
In this work, we employ additional regularization to ensure smoothness of the spatially-variable control parameter field, $\bm{\theta}$.

\subsection{Water elevation optimization} \label{subsec:elev_opt}

In this work, we minimize the model misfit with respect to water elevation observations by varying the
Manning bottom friction coefficient field.

Let $\eta_{o,i}^n$, $i=1,\ldots,B$, $n=1,\ldots,N$ denote the observation time series at $B$ tide gauges.
The model elevation field is interpolated in space at the tide gauge locations, resulting in corresponding modeled time series $\eta_{m,i}^n$. If a tide gauge lies outside the model domain, a nearest element center is used.
Time-averaged time series are denoted by $\bar{\eta}_{o,i} = \sum_n \eta_{o,i}^n/N$ and the bias-removed (centered) time series by $\hat{\eta}_{o,i}^n = \eta_{o,i}^n - \bar{\eta}_{o,i}$.
The standard deviation of observations is given by $(\sigma_{o,i})^2 = \sum_n (\hat{\eta}_{o,i}^n)^2/N$.
The Centered Root Mean Square Deviation (CRMSD) is
\begin{equation}
 \left(\mathrm{CRMSD}_i\right)^2 = \frac{1}{N} \sum_{n=1}^{N}\left( \hat{\eta}_{m,i}^n - \hat{\eta}_{o,i}^n \right)^2. \label{eq:crmse}
\end{equation}

The cost function is then defined as

\begin{equation}
 J_{o}(\mathbf{Q}, \bm{\theta}) = \frac{1}{B} \sum_{i=1}^B \frac{1}{(\sigma_{o,i})^2} \left(\mathrm{CRMSD}_i\right)^2. \label{eq:costfunc_misfit}
\end{equation}

Note that the mistfit $J_{o}$ is computed with the centered time series $\hat{\eta}_{j,i}^n, j=\{o,m\}$.
This is necessary because the model exhibits an SSH offset with respect to the observations that is not known a-priori.
The model bias $\bar{\eta}_{m,i} - \bar{\eta}_{o,i}$ can be affected by the bathymetry (defined approximately with respect to the mean sea level), river discharge, the SSH bias at the open boundary, simulated water transport between basins, and the reference level of the observations.
Similar SSH bias removal has used in previous data assimilation applications as well \cite{kurapov2011}.
Furthermore, the cost function is scaled by the inverse variance of the observations to ensure similar weight across the tidal and non-tidal tide gauges, for example.
Indeed, the standard deviation of SSH exceeds 3 m in the English channel while only being a few centimeters in the Baltic Sea.

In this application, the \emph{control variable} is the spatially varying Manning bottom friction coefficient field, i.e. $\bm{\theta}=\mu$ (see eq. (\ref{eq:bottom_friction})).
The Manning field is discretized in space with continuous linear elements (Fig. \ref{fig:elements} a).
The optimization problem is ill-posed, i.e.~there exist infinitely many Manning coefficient fields that minimize the cost function, $J_{o}$ (e.g.,~\citeA{zhang2011}).
In addition, $J_{o}$ typically exhibits multiple local minima that prevent convergence to the global optimum.
To this end, we use spatial regularization to penalize local variability of the $\bm{\theta}$ field, i.e.~we augment the cost function with a regularization term $J_{r}(\bm{\theta})$,

\begin{eqnarray}
 J(\mathbf{Q}, \bm{\theta}) &=& J_{o}(\mathbf{Q}, \bm{\theta}) + J_{r}(\bm{\theta}) \label{eq:costfunc_final}\\
 J_{r}(\bm{\theta}) &=& \alpha \| \underline{\mathbf H}(\bm{\theta}) (\Delta x)^2 \|_2^2
 =\alpha\int_\Omega \underline{\mathbf H}(\bm{\theta}):\underline{\mathbf H}(\bm{\theta}) (\Delta x)^4 \;\mathrm{d}x,
 \label{eq:costfunc_regularization}
\end{eqnarray}
\noindent where $\underline{\mathbf H}(\bm{\theta})$ denotes the Hessian of the control field with respect to the spatial coordinates, $\|\cdot\|_2$ is the $L^2$ norm, $\Delta x$ is the local mesh element size, and $\alpha$ is an unknown regularization parameter.

It should be noted that we only penalize the Hessian, i.e. the constant part and first gradients of $\bm{\theta}$ are not constrained at all.
When computing $\underline{\mathbf H}(\bm{\theta})$, we use a zero-gradient assumption for $\bm{\theta}$ at the boundary; the finite element implementation for computing $\underline{\mathbf H}(\bm{\theta})$ is presented in \ref{sec:hessian_recovery}.
The penalty term is proportional to the Frobenius norm of the $\underline{\mathbf H}(\bm{\theta})$ (sum the squares of each element).
As such, the regularization term is well-behaving in the sense that it is convex, positive definite, and invariant under coordinate system rotation.

The scaling by $\Delta x$ means that we effectively penalize the $\bm{\theta}$ variability within each element:
$\partial \bm{\theta}/\partial x \Delta x \approx \Delta \bm{\theta}$.
This regularization term ensures that the Manning coefficient field remains smooth while higher variability is allowed in regions with high mesh resolution.

The optimization problem (\ref{eq:optimization_problem}) is solved with the L-BFGS-B quasi-Newton method with bound constraints \cite{byrd1995} from the SciPy package \cite{scipy2020}.
During the iteration, we impose lower and upper bounds, $10^{-3}$ and $5^{-1}\ \mathrm{s}\ \mathrm {m}^{-1/3}$, respectively, for the Manning coefficient field; in practice the upper bound is never reached.

To facilitate applications, an inverse modeling module was as added to Thetis. The module allows defining the cost function $J_o$ based on observational time series data at station locations and the regression term $J_r$. The module also interfaces with Scipy's L-BFGS-B method and provides file IO for the control and adjoint variables.

\section{North Sea -- Baltic Sea simulation} \label{sec:baltic_configuration}

We apply the 2D shallow water model to the North Sea and Baltic Sea to simulate tidal and atmospherically-driven
water elevation dynamics.

\subsection{Model configuration}

\begin{figure}
 \centering
 \noindent\includegraphics[width=\textwidth]{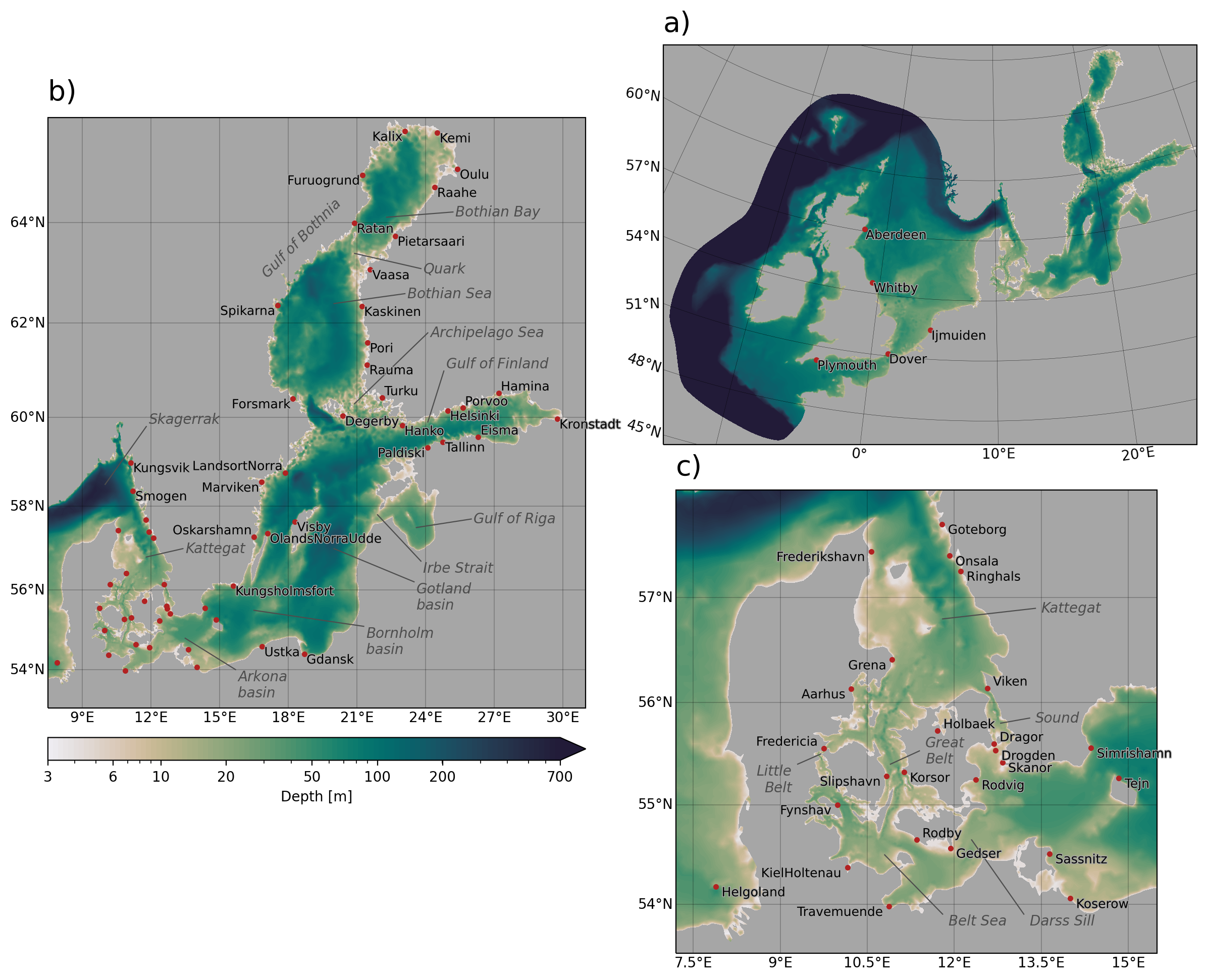}
 \caption{
 Model domain and bathymetry; (a) entire domain, (b) the Baltic Sea, and the (c) Danish Straits region.
 Red dots indicate tide gauge locations.
 }
 \label{fig:model_domain}
\end{figure}

The model domain covers the North Sea and Baltic Sea (Fig. \ref{fig:model_domain}).
The open boundary is placed beyond the continental shelf to allow reliable imposition of the tides due to greater water depth and weaker currents \cite{heemink2002,debrye2010}.
In addition, the domain is sufficiently large so that most atmospherically-driven events, such as storm surges, can be generated within the model.

\begin{figure}
 \centering
 \noindent\includegraphics[width=\textwidth]{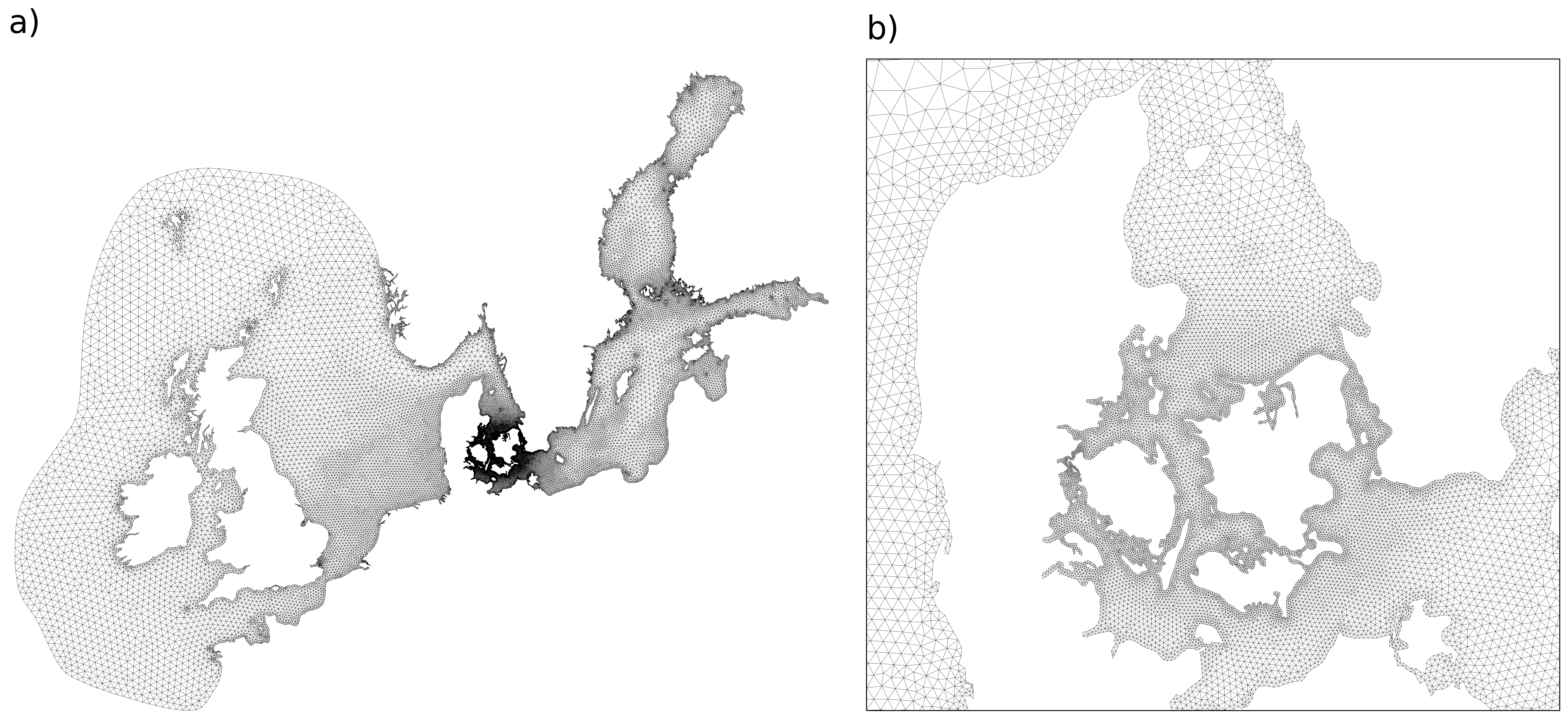}
 \caption{
 Triangular mesh of the model domain.
 The  mesh consists of 53 558 elements whose size ranges from roughly 500 m to 23 km.
 }
 \label{fig:mesh}
\end{figure}

Fig. \ref{fig:mesh} presents the triangular unstructured mesh, generated with GMSH \cite{geuzaine2009}.
We first defined a scalar field indicating the desired mesh resolution across the domain.
The coastal boundary line was extracted from the bathymetry raster at the 0.5 m contour.
In order to generate a smooth coastline, the bathymetry was first smoothed with a Gaussian filter;
stronger smoothing was applied in regions with coarser desired mesh resolution.
This procedure ensures that the coastline is compatible with the bathymetric data and also appropriate for the intended mesh resolution.
Higher mesh resolution is used along the coasts and especially in the Danish waters to better capture the complex geometry; resolution progressively decreases towards the open boundary.

Bathymetric data originates from the 1/16 arc minute EMODnet 2020 dataset \cite{emodnet2020}.
A second order P$_2$ discretization (Fig. \ref{fig:elements} c) is used for the bathymetry field, $h$, in order to improve the representation of small-scale features, such as narrow channels in the Danish Straits.
During our initial tests we noted that the geometry and bathymetry of the channels play an important role in both the volume flux between the North Sea and Baltic Sea, as well as the reflection and refraction of tides in the region.
The procedure of generating the P$_2$ bathymetry field is described in \ref{sec:p2_bathy}.
Wetting and drying is not used in the simulation as it has only a minor impact on the Baltic Sea SSH dynamics.
Minimum water depth was set to 2 m. In areas with strong tides, the minimum depth was further increased to accomodate the tidal range (based on TPXO tidal model data).

At the open boundary, the model is forced with the TPXO 9 (v1) global tidal model \cite{egbert2002} using all 15 constituents.
We impose only tidal water elevation; water velocity is relaxed towards zero using the Roe fluxes.
Our experiments indicated that including tidal velocity forcing does not improve model skill and may result in instabilities at the vicinity of the boundary.
Subtidal SSH variation is not imposed.
This configuration was found to be sufficient to represent both the mean SSH and atmospherically-driven SSH variability in the region of interest (the Danish Straits and the Baltic Sea).

A constant-in-time river discharge is imposed at 428 major rivers across the domain.
The mean river discharge was computed from the EHYPE watershed model \cite{donnelly2015} hindcast data.

Atmospheric wind stress and mean sea level pressure are imposed from the 2.5 km MetCoOp Ensemble Prediction System (MEPS)  data obtained from the Norwegian Meteorological institute.
The MEPS data set does not cover the western and southwestern parts of the model domain where
the European Centre for Medium-range Weather Forecasts (ECMWF) HRES forecast data is used instead.
The datasets are blended together in a 50 km overlapping region using a linear blending mask.
Wind stress is computed with the \cite{large2008} formulation from 10 m winds.
The atmospheric pressure and wind stress fields are discretized with linear continuous P$_1$ elements (Fig. \ref{fig:elements} a).
Viscosity was set to a constant 5 $\textrm{m}^2/\textrm{s}$ throughout the domain.

The model time step was set to 1 h.
The time step was chosen to minimize computational cost: A fully implicit solver allows longer time steps and can therefore reduce computational cost significantly \cite{karna2011}.
However, using a $\Delta t > 1\ \textrm{h}$ resulted in slower convergence of the solver and thus higher overall cost.
Using 1 h time step for modeling tidal dynamics is appropriate as we are using a second-order implicit Runge-Kutta solver which can represent nonlinear processes accurately. In addition, during the Runge-Kutta sub-iteration, all forcing fields are evaluated twice in each time step which, again, increases the accuracy.
Our preliminary tests did not show any significant deterioration of the SSH performance with 1 h time step (not shown).
We did notice, however, that the commonly-used, asymptotically unstable Crank-Nicolson time integrator did exhibit numerical instability manifested in a noisier velocity field.

The modeled SSH is compared against tide gauge observations at 56 tide gauges across the model domain (Fig. \ref{fig:model_domain}).
The observational data was obtained from the Copernicus Marine Service (CMEMS) catalog \cite{cmems-baltic,cmems-northsea}.
Only tide gauges that had nearly complete data coverage over the modeled period were included.
The tide gauge time series were manually quality-checked to remove spurious SSH values (e.g., spikes).

\subsection{Simulation period and initial condition}

The Manning coefficient is optimized for a 16.5 day period, spanning from June 1 00:00 UTC to June 17 12:00 UTC, 2019.
To exclude initial adjustment effects from the optimization, the cost function is not evaluated during the
first 2.5 days, i.e.~the model misfit is calculated over a 14-day period.
This period is henceforth called the optimization period.

We chose to use a 14-day period to cover a full spring-neap tidal cycle as well as several seiche waves in the Baltic Sea.
Early experiments showed that a shorter, 2-day period results in over-fitting as the model adjusts to particular realizations of coastal waves.
Generally, choosing a suitable optimization window length is a trade-off, as the computational cost is directly proportional to the number of time steps.

On June 1, 2019, the model is initialized from a spun-up initial state.
Using a realistic initial condition is important for the optimization process.
According to our sensitivity runs, the mean SSH in the Baltic Sea reaches an equilibrium in roughly 1 month (not shown)
and this transient adjustment must be excluded from the optimization process as it would skew the results.
Moreover, a relatively good guess for the  Manning field is needed for the spin-up run as bottom friction affects the volume flux to/from the Baltic Sea and therefore the mean SSH.
The initial condition was generated as follows:
First, 10 iterations of the optimization process are carried out starting the model from rest (zero initial water elevation and velocity; the Manning coefficient was set to 0.03 $\mathrm{s}\ \mathrm {m}^{-1/3}$). As a result, a first guess of a suitable Manning coefficient field, $\mu_0$, is obtained.
Using the $\mu_0$ field, a 1-month spin-up run (May 1, 2019 to June 1, 2019) is then carried out to generate an initial condition for June 1. The spin-up run used the ERA5 atmospheric reanalysis data as forcing \cite{era5}.

To initialize the final optimization, the Manning coefficient is set to 0.03 $\mathrm{s}\ \mathrm {m}^{-1/3}$ everywhere in the model domain.
The optimization was run for 40 iterations, after which the changes in the friction field were small.

Once the optimal Manning coefficient field has been found, the results are validated with a 3-month simulation ranging from June 1 to August 31, 2019.
The start date is the same as with the optimization period and the same model initial condition is used for elevation and velocity.
The first 2.5 days are again excluded from the analysis.
The longer validation run allows to verify that the Manning coefficient has not been tuned to fit particular events during the optimization period (over-fitting), i.e., the optimized model is able to represent SSH dynamics accurately in general.

\subsection{Performance metrics}

The model skill is quantified with standard statistical metrics.
Using the notation from Section \ref{subsec:elev_opt} (we drop the station index $i$ for brevity),
the bias and correlation coefficient ($R$) are defined as:
\begin{eqnarray}
 \mathrm{BIAS} &=& \bar{\eta}_{m} - \bar{\eta}_{o}, \\
 R &=& \frac{1}{\sigma_o\sigma_m}\frac{1}{N}\sum_{n=1}^{N} \hat{\eta}_{m}^n \hat{\eta}_{o}^n.
\end{eqnarray}

CRMSD is related to $\sigma_m$ and $R$ through the equation,
\begin{eqnarray}
 \mathrm{CRMSD}^2 &=& \sigma_o^2 + \sigma_m^2 - \sigma_o\sigma_m R, \label{eq:taylor_diag}
\end{eqnarray}
which can be visualized in a Taylor diagram \cite{taylor2001}.
In this work, we are normalizing the Taylor diagram by scaling the variables with $\sigma_o$:
\begin{eqnarray}
 \mathrm{NCRMSD}^2 &=& 1 + \sigma_m'^2 - \sigma_m' R,\label{eq:taylor_diag_normalized}\\
 \mathrm{NCRMSD} &=& \frac{1}{\sigma_o}\mathrm{CRMSD},\\
 \sigma_m' &=& \frac{\sigma_m}{\sigma_o},
\end{eqnarray}
where $\mathrm{NCRMSD}$ and $\sigma_m'$ are the normalized CRMSD and standard deviation of the model, respectively.
Normalization leads to dimensionless metrics and permits the comparison of different data sets.

\section{Results} \label{sec:results}

\subsection{Optimization}

\begin{figure}
 \centering
 \noindent\includegraphics[width=0.5\textwidth]{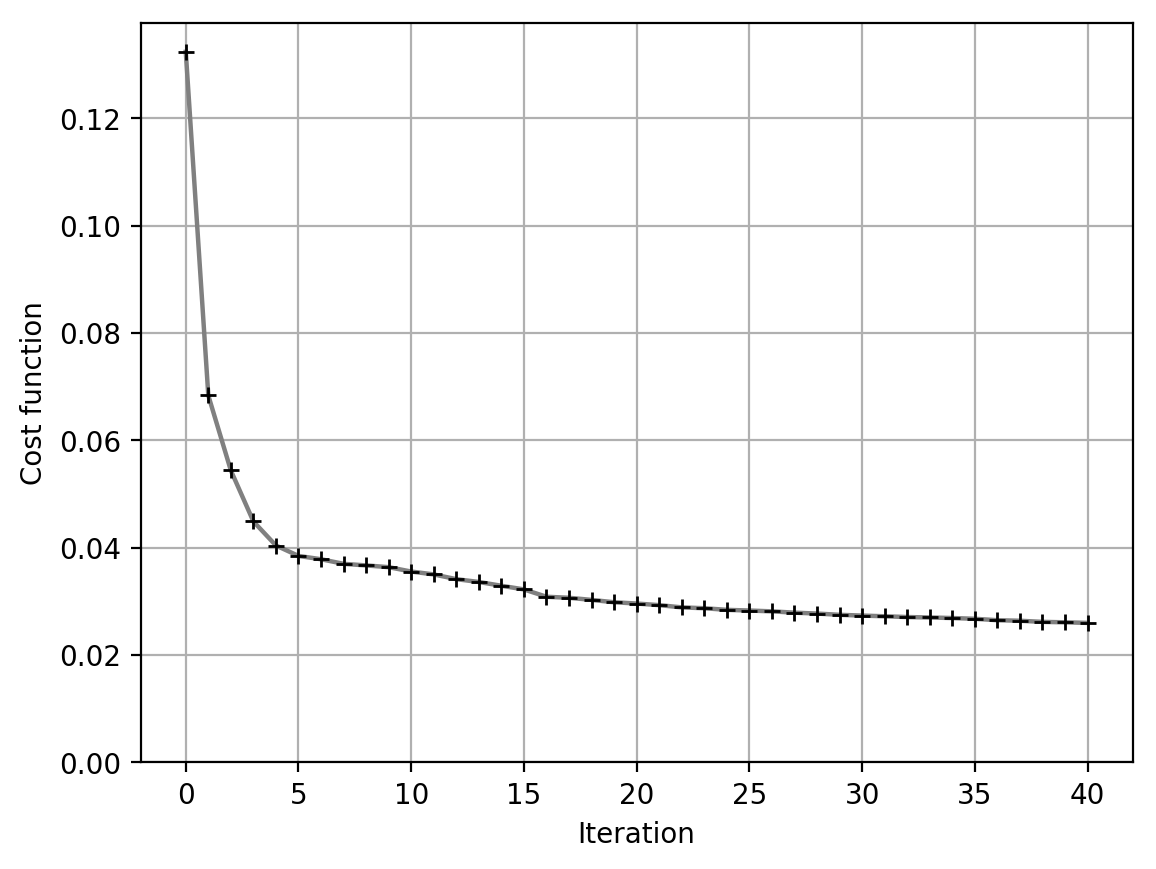}
 \caption{
 Evolution of the cost function during the optimization iteration.
 }
 \label{fig:optimization_costfunc}
\end{figure}

The evolution of the cost function during the optimization is shown in Fig. \ref{fig:optimization_costfunc}.
During the first 5 iterations the cost function decreases rapidly and begins to stagnate between iterations 20 and 40.
The regularization parameter, $\alpha$, was set to 400 as it appeared to result in a smooth Manning coefficient field without constraining the optimization too much.

\begin{figure}
 \centering
 \noindent\includegraphics[width=\textwidth]{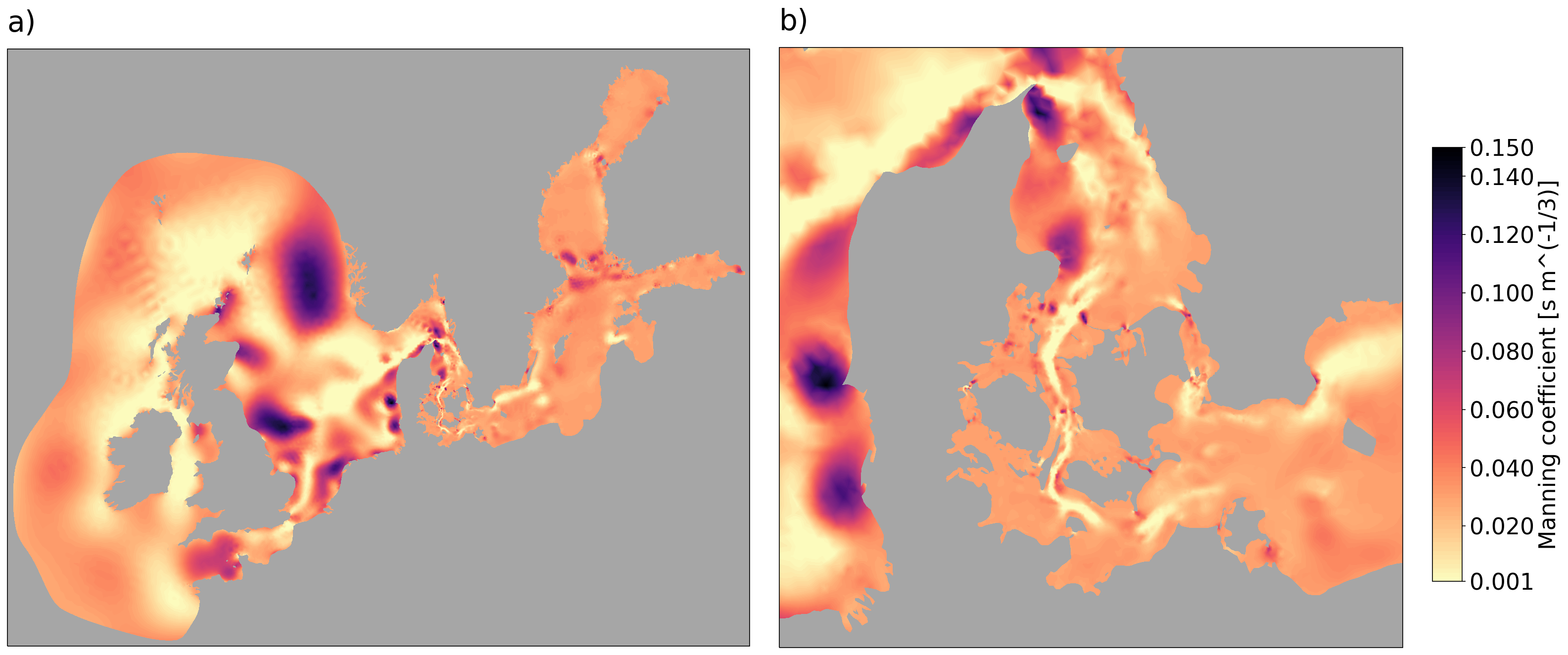}
 \caption{
 Optimized Manning coefficient field after 40 iterations; (a) full domain; (b) close-up on the Danish Straits region.
 }
 \label{fig:map_manning}
\end{figure}

The final optimized Manning coefficient field is shown in Fig. \ref{fig:map_manning}.
The Manning field shows significant spatial variability throughout the model domain, i.e.~not only in the vicinity of the tide gauge locations. This suggests that the optimization process alters the propagation of long surface waves instead of finding artificial local minima at the observation locations.
Fig. \ref{fig:map_manning} shows some oscillatory patterns in the Atlantic Ocean north of Scotland. These patterns are due to the spatial regularization method which does not ensure a completely smooth Manning coefficient field. In practice, however, the effect of these oscillations is small.

In the North Sea and around the British Isles, the friction coefficient is altered to improve the propagation and reflection of the tides;
In the Danish waters (Fig. \ref{fig:map_manning} b), friction is lowered in the Great Belt and across the Darss Sill.
This is the dominant route for water transport between the Baltic Sea and the North Sea \cite{mohrholz2015,grawe2015,mohrholz2018} and the friction coefficient has a great impact on the barotropic volume flux, driven by water elevation difference across the Danish Straits.

In the Baltic Sea, the friction coefficient is close to the initial value especially in the deeper parts of the basin. In these regions (and also close to the open boundary) the gradient of the cost function shows very small values over the entire optimization process (not shown), suggesting that the cost function is not very sensitive to bottom friction in these regions.
However, the Manning coefficient is altered at certain shallow regions, such as the Quark, Archipelago Sea and the Irbe Strait. These locations tend to have stronger depth-averaged currents and control the volume transport to the Bothnian Bay, Bothnian Sea, and Gulf of Riga, respectively.
The Archipelago Sea is largely under-resolved with the used mesh resolution which plausibly explains the need for higher friction.
\citeA{massmann2010} also found that adjoint-based sensitivity indicates higher bottom friction in the vicinity of unresolved islands.
Overall, the optimized Manning coefficient field is complex and would be difficult to find through manual manipulation or brute force methods.

\begin{figure}
 \centering
 \noindent\includegraphics[width=0.9\textwidth]{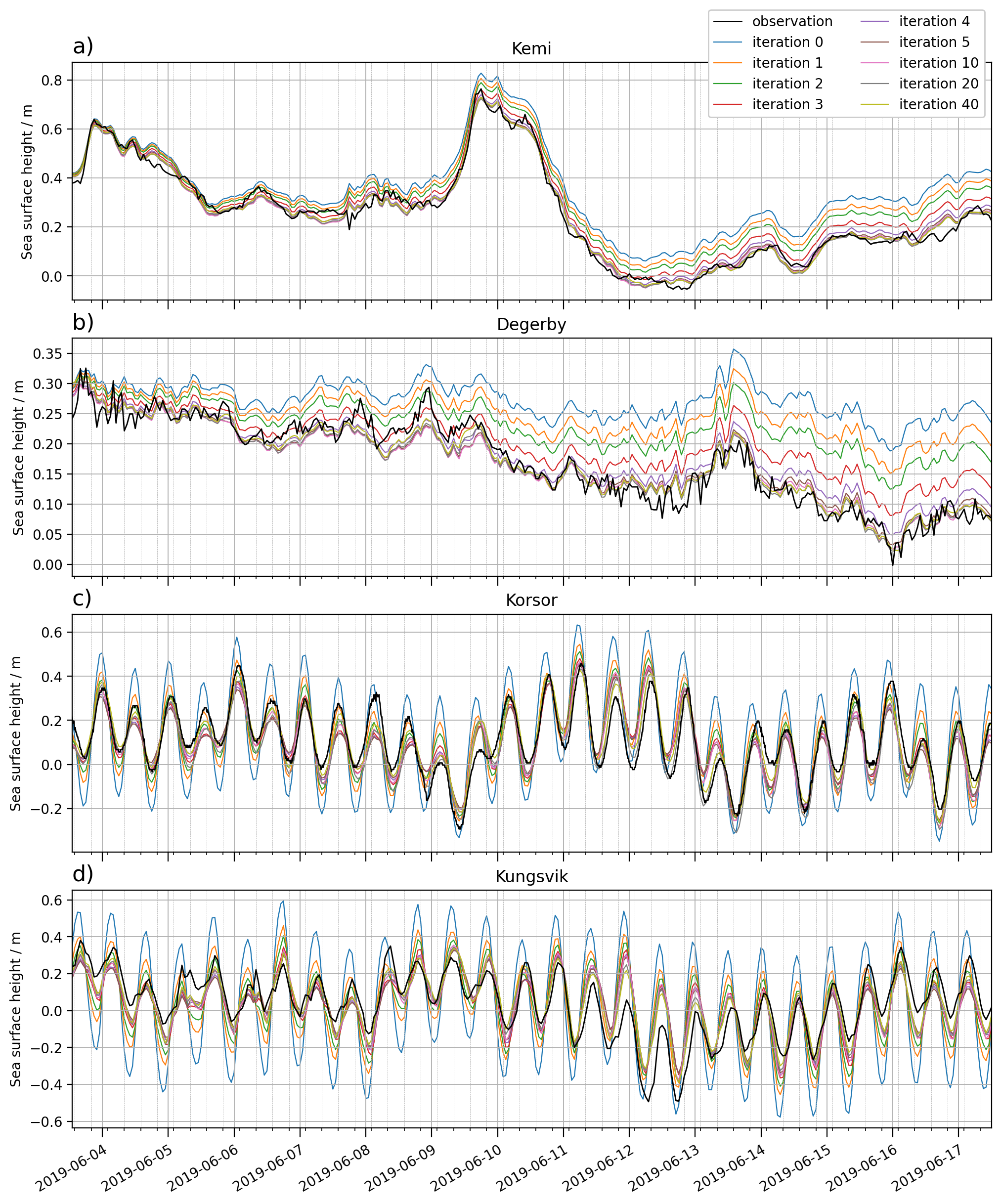}
 \caption{
 Example time series during the optimization.
 The black line stands for the observation; colored lines indicate model results at different stages of the optimization progress; iteration 0 indicates the initial guess.
 The data sets have been bias corrected for visual comparison.
 }
 \label{fig:optimization_timeseries}
\end{figure}

Example time series of the optimization process are shown in Fig. \ref{fig:optimization_timeseries}.
In the Baltic Sea (panels a and b), the optimization mainly affects the slowly varying mean water elevation.
This is related to the volumetric transport into the Baltic Sea.
The Degerby station lies close to a nodal point of the basin's seiche oscillations and thus it is often regarded as a metric of the total water volume in the basin.
Panel (b) thus suggests that the emptying of the basin is initially underestimated;
after 5 iterations, the model already tracks the observations fairly closely.

Panels (c) and (d) show examples of tidal stations in the Great Belt and Skagerrak, respectively.
In both cases, the tidal range is initially overestimated but the model recovers a reasonable tidal range as the iteration progresses.
The optimization also affects the phase of the tides to some extent.
This is plausibly achieved by altering the relative strength of the tidal waves that propagate though the various parallel straits, for example.
Based on our experience, these kinds of effects are difficult to achieve through manual manipulation of the Manning coefficient field.
It should be noted that in the Danish Straits the solution converges slower compared to the Baltic stations (a and b) and the tidal waves are not fully replicated even after 40 iterations.

\begin{figure}
 \centering
 \noindent\includegraphics[width=1.1\textwidth]{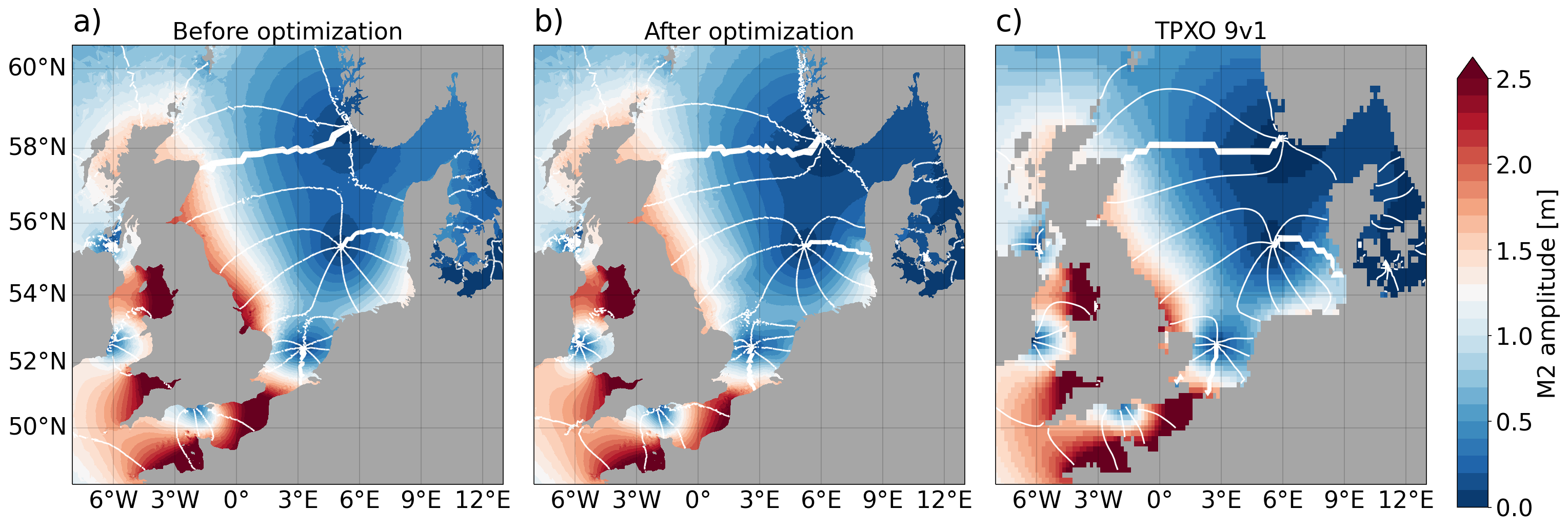}
 \caption{
 Amplitude and phase of the M2 tidal constituent estimated for the validation period. (a) uncalibrated model; (b) model after optimization; (c) TPXO solution. White lines denote the phase with 45$^\circ$ intervals; the thick line is the 0$^\circ$ isocontour.
 }
 \label{fig:optimization_M2}
\end{figure}

Figure \ref{fig:optimization_M2} shows the estimated amplitude and phase of the M2 tidal constituent before and after the optimization (panels (a) and (b), respectively).
Overall, the M2 tidal pattern is similar in both cases, but there are some significant differences.
Before optimization, the model overestimates the M2 amplitude by several tens of centimeters in the west coast of Denmark and in the Danish Straits.
Thus the overestimation of tidal range shown in Fig. \ref{fig:optimization_timeseries} is a generic feature of the entire region.
Optimization also affects the phase and moves the amphidromic points slightly.
After optimization, the M2 amplitude and phase are in good agreement with the TPXO model (panel (c)).

\subsection{Validation}

\begin{figure}
 \centering
 \noindent\includegraphics[width=0.85\textwidth]{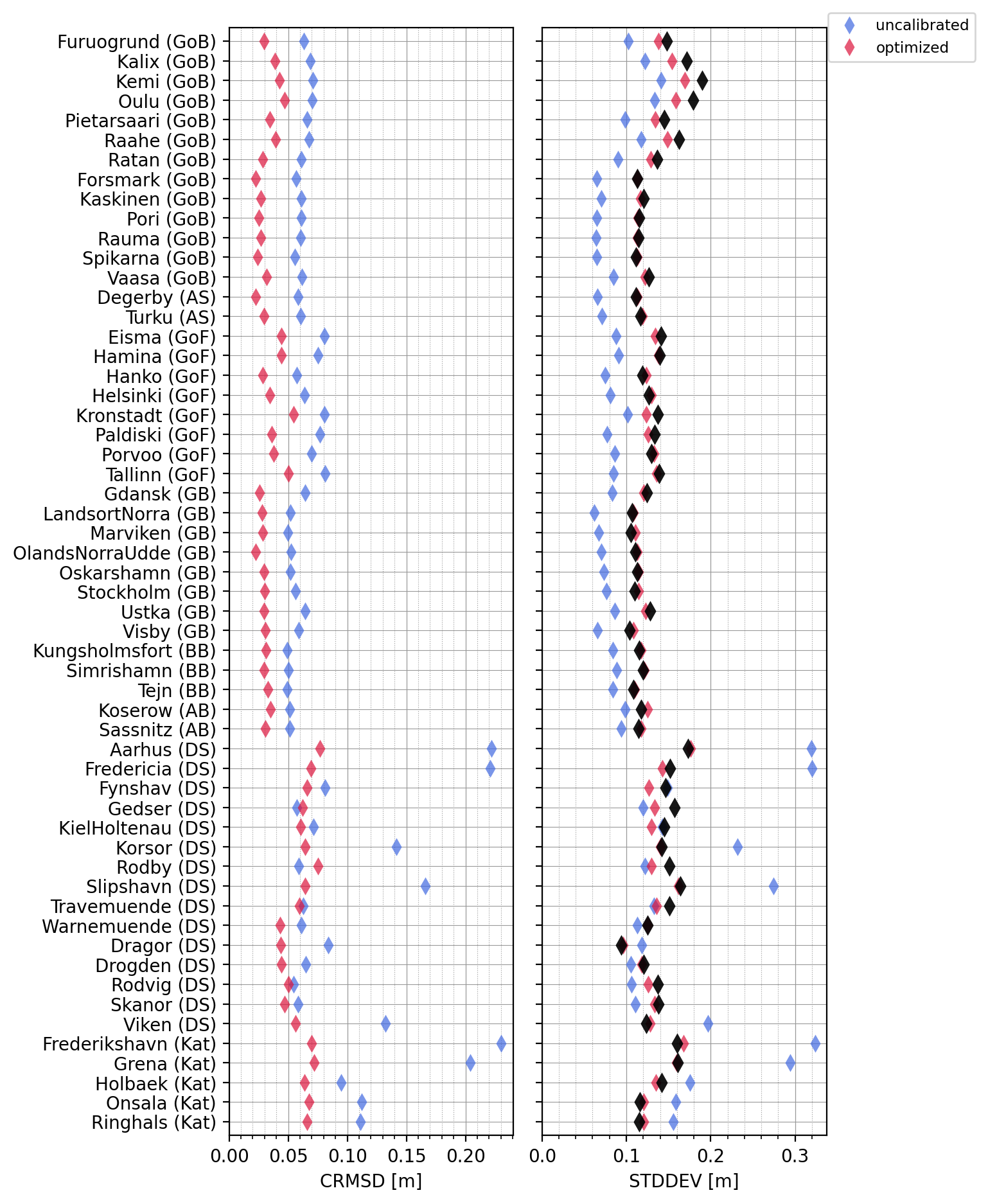}
 \caption{
 Performance metrics for each tide gauge for the validation period (June 1 to August 31, 2019).
 The black diamonds denote the standard deviation of the observations.
 Sub-basins are indicated by the abbreviations: GoB, Gulf of Bothnia; AS, Archipelago Sea; GoF, Gulf of Finland; GB, Gotland Basin; BB, Bornholm Basin; AB, Arkona Basin; DS, Danish Straits; Kat, Kattegat.
 }
 \label{fig:tg_validation_stats}
\end{figure}

Model performance metrics for the validation period are presented in Fig. \ref{fig:tg_validation_stats}.
The blue and red markers indicate the initial and optimized model, respectively.
The optimization improves the model performance at all stations.
In the Baltic Sea, the CRMSD improves from 5 to 8 cm range to 2 to 6 cm range;
In many locations in the Gotland Basin and the Bothnian Sea, the optimized model reaches CRMSD of 3 cm or less.
Slightly higher CRMSD is observed in the Bothnian Bay and Gulf of Finland.
In the Danish Straits, the deviation is higher, the optimized model's CRMSD ranging between 4 and 8 cm.
In this case, the improvement is substantial:
At certain locations (e.g.~Aarhus, Fredericia, Frederikshavn) the CRMSD is reduced by nearly a factor of 3.
This is due to the complex reflection and interaction of tidal waves in this region which is difficult to replicate without careful tuning.
For example, the tides propagate through the straits with different phase speeds, resulting in a complex superpositioning of tidal waves in the Great Belt; our initial tests indicate that a sufficiently good representation of the geometry is essential for capturing these processes.

The second panel in Fig. \ref{fig:tg_validation_stats} shows the standard deviation of the observed and modeled and time series;
The black markers indicate the observations.
Also here, the optimization clearly improves the performance.
The model has a slight tendency to underestimate variability at certain stations in the Gulf of Bothnia and Danish Straits.

\begin{figure}
 \centering
 \noindent\includegraphics[width=0.7\textwidth]{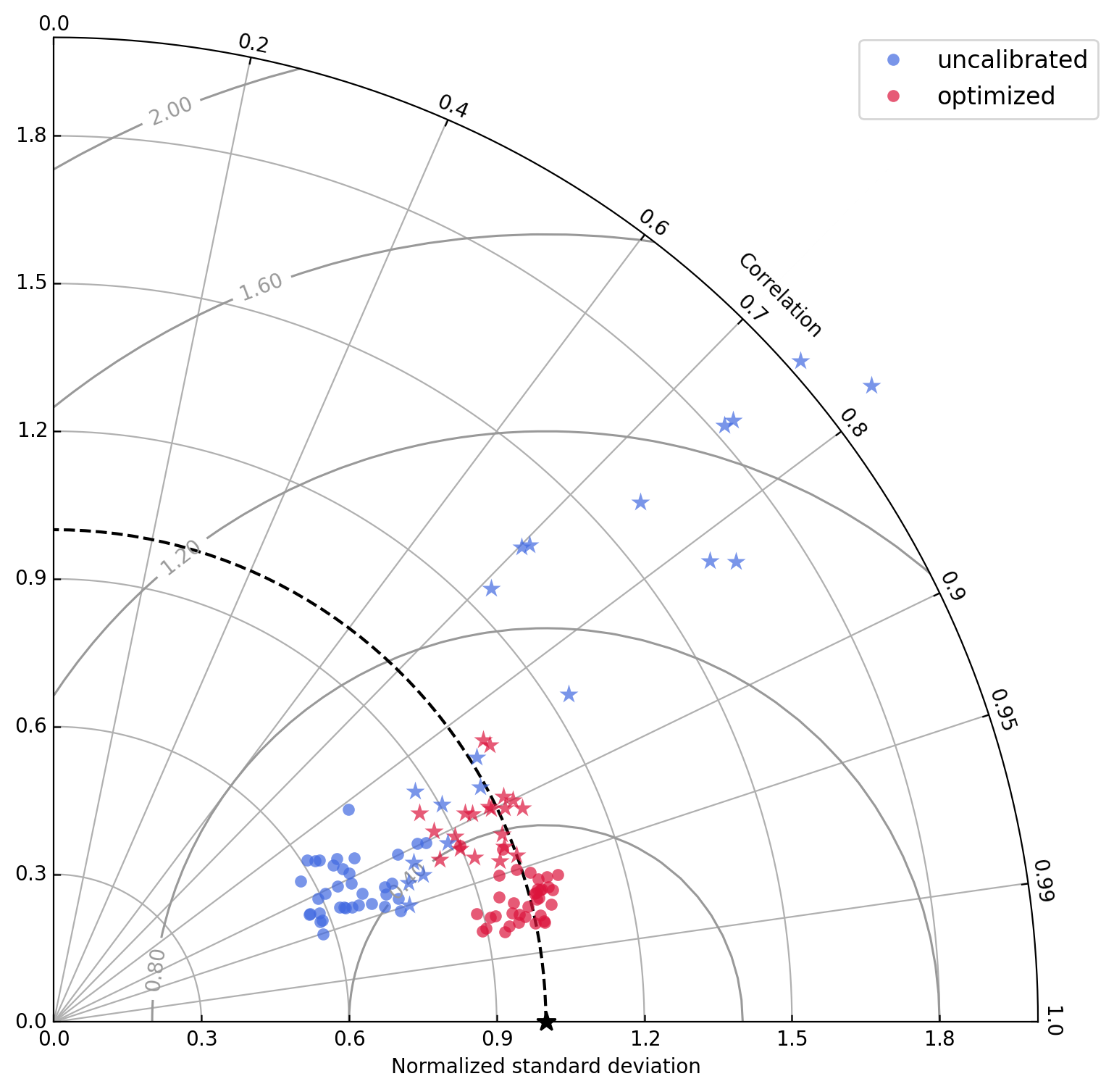}
 \caption{
 Taylor diagram of tide gauge data for the validation period.
 The round and star symbols indicate stations in the Baltic Sea and Danish waters (Danish Straits and Kattegat), respectively.
 }
 \label{fig:tg_validation_taylor}
\end{figure}

Fig. \ref{fig:tg_validation_taylor} presents a Taylor diagram for the validation data.
Also here the improvement in model skill is clear.
In the Baltic Sea (round markers) the optimized model performs very well;
at most stations the correlation coefficient is above 0.95 and standard deviation is close to the observations.
In the Danish waters (star markers), the correlation coefficient is lower but still above 0.83 in all cases.

\begin{figure}
 \centering
 \noindent\includegraphics[width=0.8\textwidth]{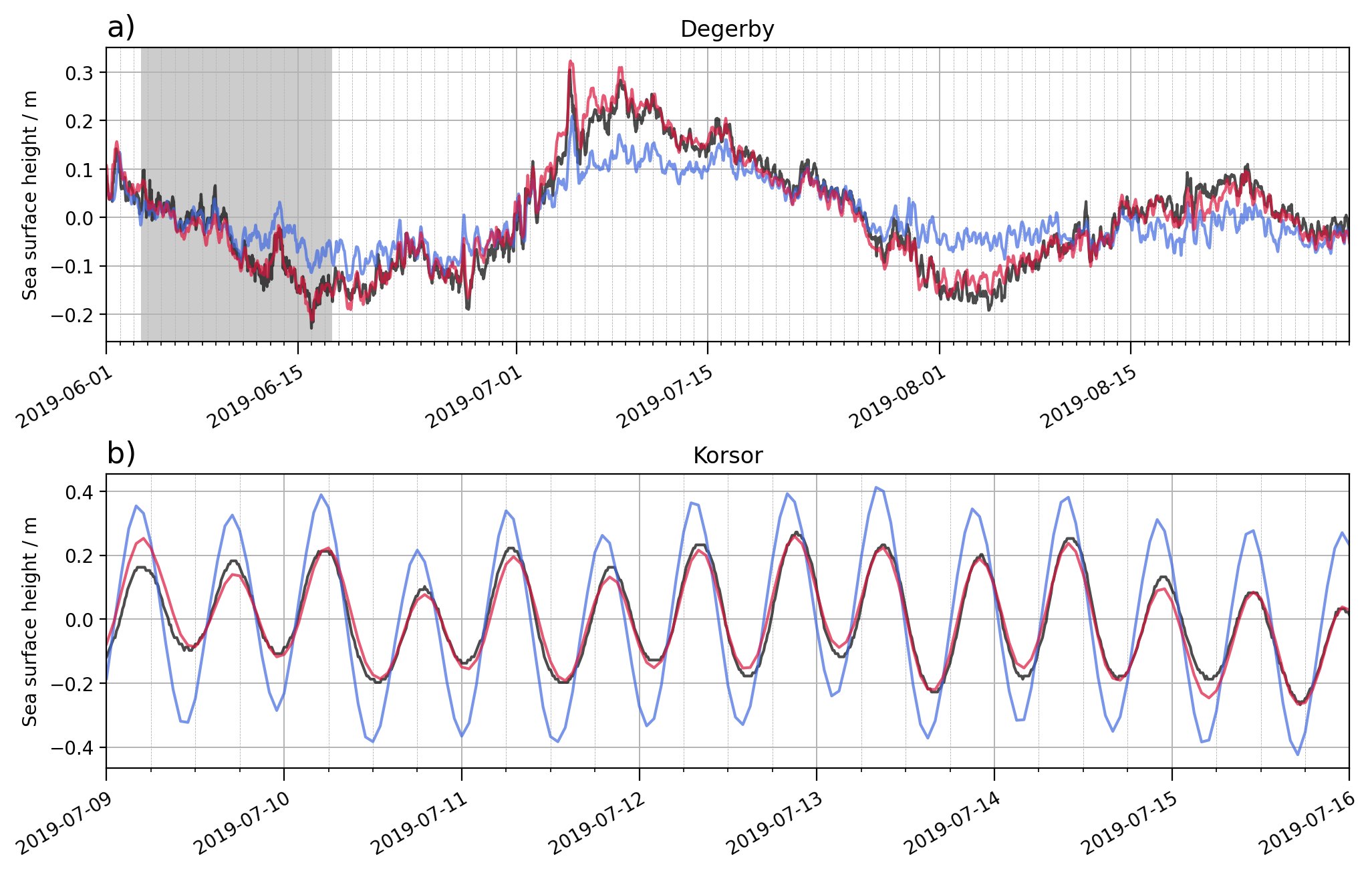}
 \caption{
 Example time series from the validation run at the (a) Degerby and (b) Korsor stations.
 Panel (a) covers the entire validation period; gray shading indicates the 14-day optimization period.
 Panel (b) shows tidal variability in the Great Belt region.
 The black, blue and red lines stand for the observations,
 uncalibrated, and optimized model, respectively.
 The data sets have been bias corrected for visual comparison.
 }
 \label{fig:validation_timeseries}
\end{figure}

Two time series examples are shown in Fig. \ref{fig:validation_timeseries}.
Panel (a) shows water elevation at Degerby station for the entire validation period.
Similarly to Fig. \ref{fig:optimization_timeseries}, the uncalibrated model tends to underestimate the slowly-varying SSH changes, suggesting that the transport through the Danish Straits is underestimated.
The optimized model, on the other hand, tracks the volumetric changes quite accurately.
Panel (b) shows tidal SSH signal at Korsor station in the Great Belt for a 7-day period outside the optimization window.
Also in this case, the optimized model produces both a realistic tidal range as well as sub-tidal variability.

\subsection{Computational cost}

The simulations were carried out on AMD Rome 7H12 CPUs on the CSC Mahti cluster using 8 MPI processes.
The forward model runs roughly 1800 times faster than real time: a 1-day and a 1-month simulation take roughly 48 s and 24 min, respectively.
Evaluating the gradient of $J$ (including both the forward and adjoint solves) in the 16.5 day optimization period takes roughly 45 min, i.e.~about 3.4 times longer than running the forward model alone.
In each L-BFGS-B iteration typically only one, and at most two, gradient evaluations are needed.
Running the entire optimization for 40 iterations took 36 h.

Comparing the cost against other adjoint model implementations is not straightforward as computational cost is rarely reported.
\citeA{yaremchuk2016} used a manually-coded adjoint of the Navy Coastal Ocean Model (NCOM); a single tangent linear model (TLM) and adjoint evaluation costs approximately 11 times as much as the forward model.
\citeA{villaret2016} used an AD-generated TLM to study the sensitivity of the Telemac2D/Sisyphe morphodynamic model. They report that one TLM evaluation costs about 3 times as much as the forward model.
Considering that a TLM evaluation is cheaper than the full forward-backward solve, this suggests that our adjoint implementation is at least as efficient as the implementations referenced above.

\section{Discussion} \label{sec:discussion}

We have presented an adjoint-based optimization of a 2D water elevation model.
One of our main findings is that for the optimization process to work, the model misfit must be dominated by the control variable, i.e.~bottom friction in this case.
Otherwise, the optimization may end up compensating errors due to other sources, e.g.,~by generating artificial friction patterns near the boundary in the case of poor boundary forcings.
Consequently, one needs a sufficiently good mesh, bathymetry, boundary conditions and atmospheric forcing.
The open boundaries should be located sufficiently far away from the domain of interest.
In addition, a realistic initial condition is crucial for robust optimization.

In this work we are using a 2D shallow water model.
As such, our model does not include water density effects, or baroclinic processes. Wind wave effects are also not included.
Water density difference between the North Sea and the northern Baltic Sea introduces a steady elevation gradient across the Baltic basin.
Baroclinic processes can be important especially in the Danish Straits where density gradients are strong and fluctuate with in- and outflow events.
The freshwater outflow in Kattegat/Skagerrak can exhibit baroclinic eddies, which affect local water elevation.
Wind wave effects are most pronounced under storm conditions and can enhance storm surges via Stokes drift or wave build-up near the coast \cite{kanarik2021}.
The results indicate, however, that a 2D barotropic model can replicate the majority of the SSH variability with good accuracy.

The presented inverse modeling procedure resembles the so-called four-dimensional variational assimilation (4D-Var) method  widely used in both atmospheric \cite{ledimet1986} and oceanic \cite{thacker1988,wunsch2007} applications.
Compared to simpler methods (e.g.~3D-Var), the time dependency is consistently treated with the forward-backward solution procedure.
In addition, the constraint (\ref{eq:forward}) ensures that the solution always satisfies the model equations which is not the case in many widely-used data-assimilation methods (e.g. simple nudging of the model state, and filtering methods including the Kalman filter and its approximations).
One notable difference is that some data-assimilation implementations use the ensemble method to estimate the gradient of the cost function, which is generally more expensive than solving the adjoint equation as the cost is directly proportional to the size of the ensemble.

One common pitfall in gradient-based optimization is over-fitting which must be controlled with some form of regularization.
\citeA{heemink2002} and \citeA{zhang2011} reduced the dimensionality of the friction field considering only a small subset of the grid's nodal values.
In this work, in addition to using a sufficiently long optimization period, we included a regularization term to penalize the second derivative of the control variable.
This procedure is sufficient to avoid strong ``bipolar'' friction adjustment at the tide gauge locations.
The final Manning coefficient field is complex and highly variable across the domain; friction is altered also in regions where no observation data is used (e.g.,~around the British Isles and the Irish Sea).
It is worth noting, however, that shallow water surface waves are typically relatively long and smooth which also reduces the risk of spatial over-fitting.

The presented optimization process is robust as most tests resulted in a similar friction field.
We did observe some dependency on the mesh, however.
As such, the optimized friction field should be regarded as a model and mesh dependent parameter with limited physical interpretability (e.g.~with respect to sea bed roughness).
Indeed, capturing SSH dynamics requires suitable dissipation at the right locations and the model's dissipation characteristics are governed by the discretization, mesh resolution as well as physical parametrizations.

The fact that the model skill is lower in the Danish waters compared to the Baltic Sea stations is consistent with other modeling results.
Compared to the 1 nautical mile Nemo-Nordic 2.0 3D baroclinic model \cite{karna2021}, the presented results are comparable or better.
Although the simulation period, model configuration, and forcings are different, \citeA{karna2021} state similar model skill in the Baltic Sea and Danish waters (CRMSD below 7 cm and 10 cm, respectively).
However, at certain locations, e.g.~Fredericia, accuracy is lower, CRMSD being above 12 cm.
This is most likely due to the fact that the complex geometry of the Danish Straits cannot be properly represented with a 1 nautical mile (1.8 km) structured grid.

As tuning of bottom friction is necessary in many coastal applications,
the presented optimization procedure is a promising step towards automated calibration of ocean models.
Manual tuning of spatially variable friction coefficient can be a very time consuming task.
Without an adjoint model, the options for automated tuning are limited: one has to rely on a brute-force search, or ensemble-based estimates of the cost function gradient, for example.
In both approaches, computational cost is proportional to the degrees of freedom (number of nodes) of the control field.
In this application, one iteration takes 3.4 times longer than running the forward model, i.e.~the cost is roughly equivalent to running a 3-member ensemble.
The accuracy, however, is much better as the gradient ${\mathrm dJ}/{\mathrm d\bm{\theta}}$ is computed exactly.

Adjoint-based optimization has many similarities with machine learning methods \cite{sonnewald2021}.
The backward-in-time adjoint iteration resembles the backpropagation training methods.
In addition, regularization is a key ingredient to avoid over-fitting and a separate validation set is commonly used to inform the final model selection.
In this application, the cost function is defined as the centralized root mean square deviation, scaled by the observation variance.
This is in fact equivalent to whitening the target signal (i.e.~removing the bias and scaling to unit variance), commonly used as a pre-processing step in machine learning.
In contrast to data-driven machine learning methods, the benefit of adjoint-based physical modeling is that the results are guaranteed to be physically sound, i.e.~satisfy the physical principles encoded in the forward equations.

\section{Conclusions} \label{sec:conclusions}

We have presented an adjoint-based bottom friction optimization in a 2D water elevation model for the Baltic Sea.
The discrete adjoint model is automatically generated from the symbolic forward model equations.
The Manning coefficient optimization procedure is robust and yields good, generalizable results.
Achieving similar performance via manual tuning would be challenging.
The model performs well, especially in the complex Danish Straits region that is difficult to model with structured grid models (e.g., \citeA{karna2021}), highlighting the benefit of variable-resolution unstructured meshes.

Leveraging the symbolic representation of the forward model, the adjoint model can be generated and solved with very few user efforts:
The user only needs to implement the forward model and the desired cost function.
The resulting adjoint model is computationally efficient: evaluating the gradient of the cost function takes 3.4 times as long as running the forward model alone.
This work demonstrates that domain specific language frameworks, based on high-level abstractions and code generation, can facilitate automated model analysis and provide easy access to generic inverse modeling capability.

\appendix

\section{Quadratic bathymetry field} \label{sec:p2_bathy}

As the coastal topography exerts a leading-order control on water circulation, it is crucial to obtain an accurate numerical representation of the bathymetry.
Typically the accuracy of the bathymetric field is restricted by the mesh resolution.
In this work, we discretize the bathymetry field with a second order P$_2$ element (Fig. \ref{fig:elements} c) to increase the intrinsic resolution.

A common bottleneck with order $p>1$ polynomials is the loss of monotonicity near sharp gradients: The polynomial representation can lead to large oscillations that exceed the bounds of the original data. Water depth may become negative in the vicinity of steep slopes, for example.

We generate a smooth P$_2$ bathymetry field as follows:
First, we generate a refined triangular mesh by splitting each triangle uniformly into four (i.e.~joining the P$_2$ element nodes).
On the refined mesh, we define a preliminary P$_1$ bathymetry field and interpolate the raw raster bathymetry data on it.
This field does not exhibit any spurious overshoots due to the linear basis functions.
The preliminary bathymetry field is then smoothed with Laplacian diffusion.
Given an preliminary bathymetry field $h^{(1)}$ and a test function $\phi$ in the P$_1$ space, the smoothing operator is

\begin{eqnarray}
r &=& \frac{ ( | \nabla h^{(n)} | \Delta x )^\gamma }{2 (h^{(n)})^\beta } \\
\int_\Omega (h^{(n+1)} - h^{(n)}) \phi dx &=& - \int_\Omega \alpha r \nabla h^{(n)} \cdot \nabla \phi dx, \quad \forall \phi
\end{eqnarray}
\noindent with $\alpha=20$, $\gamma=5/2$, and $\beta=1/2$.
On the right hand side, the Laplacian diffusion operator has been integrated by parts, omitting the interface terms.
The diffusion coefficient, $r$, depends on the local bathymetry change in each element.
It resembles the dimensionless bathymetry slope metric commonly used in sigma coordinate models (recovered by setting $\gamma=\beta=1$; \citeA{haney1991,mellor1998}).

\begin{figure}
 \centering
 \noindent\includegraphics[width=0.95\textwidth]{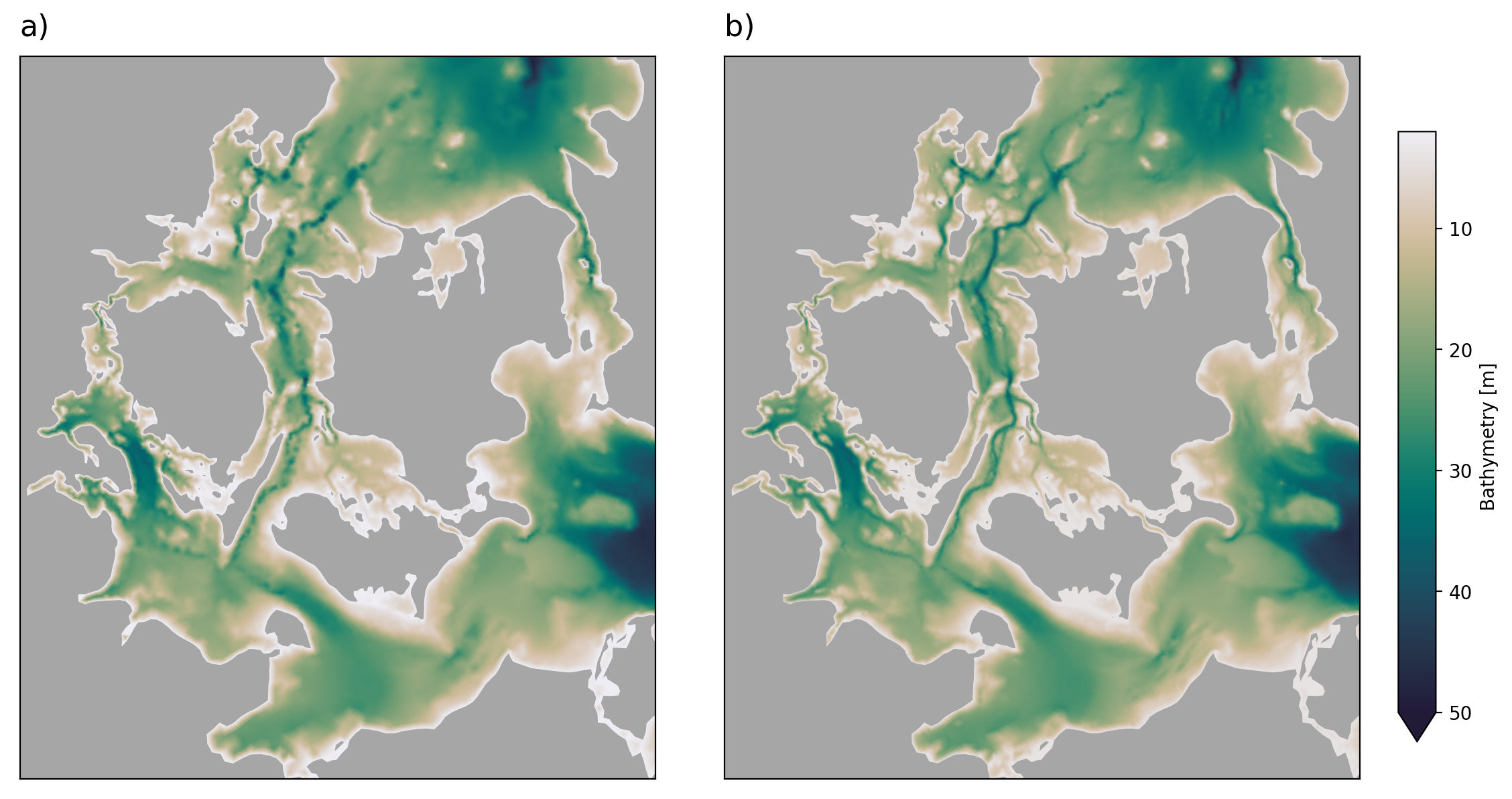}
 \caption{
 Comparison of bathymetry fields in the Danish Straits region, (a) the raster bathymetry data interpolated onto the P$_1$ elements; (b) the final P$_2$ bathymetry field.
 }
 \label{fig:bathy_comparison}
\end{figure}

The smoother is applied 30 times. We then use $L^2$ projection to cast the smoothed preliminary bathymetry on a P$_2$ field on the original mesh. Although the procedure is not guaranteed to be monotonic, the smoothing step essentially ensures that the projection does not generate severe overshoots.
A comparison of the final P$_2$ bathymetry and bathymetry interpolated on the P$_1$ elements on the original mesh are shown in Fig. \ref{fig:bathy_comparison}.
The P$_2$ discretization allows more realistic representation of the narrow channels in the Danish Straits.
In the P$_1$ case, the channels are not entirely continuous which affects volume transport.

\section{Hessian recovery procedure} \label{sec:hessian_recovery}

The cost function described in Section \ref{subsec:elev_opt} involves a regularization
term that depends on the Hessian of the control field with respect to the spatial coordinates.
In practice, the control field is discretized using linear finite elements, which are not
twice continuously differentiable.
In order to approximate the second spatial derivatives of the (continuous) control field, we apply
an $L^2$ projection recovery procedure.
The gradient of the discrete field is projected into $L^2$ space.
A second application then obtains the Hessian approximation.
We opt to perform the Hessian recovery in P$_1$ spaces and apply the two recovery steps simultaneously using a mixed finite element method.

Let $\theta^h$ denote the finite element approximation of the control field and
$\mathbf g^h(\theta)$ and $\underline{\mathbf H}^h(\theta)$ denote the corresponding approximations of its
gradient and Hessian, the latter of which we seek to obtain.
Let $\boldsymbol\phi$ and $\underline{\boldsymbol\phi}$ denote test functions in vector and
tensor P$_1$ spaces, respectively.
The mixed formulation may then be written as

\begin{eqnarray}
  \label{eq:hessian_recovery1}
  \int_\Omega \mathbf g^h \cdot \boldsymbol\phi \;\mathrm dx
  = -\int_\Omega \theta^h (\nabla\cdot\boldsymbol\phi) \;\mathrm dx
  + \int_\Gamma \theta^h (\boldsymbol\phi\cdot\mathbf n) \;\mathrm ds
  + \int_{\mathcal I} \{\!\!\{\theta^h\}\!\!\} {[}\!{[}\boldsymbol\phi\cdot\mathbf n{]}\!{]} \;\mathrm dS
  \quad \forall\boldsymbol\phi\\
  \label{eq:hessian_recovery2}
  \int_\Omega \underline{\mathbf H}^h:\underline{\boldsymbol\phi} \;\mathrm dx
  = -\int_\Omega \mathbf g^h \cdot (\nabla\cdot\underline{\boldsymbol\phi}) \;\mathrm dx
  + \int_\Gamma \mathbf g^h \cdot (\underline{\boldsymbol\phi}\cdot\mathbf n) \;\mathrm ds
  + \int_{\mathcal I} \{\!\!\{\mathbf g^h\}\!\!\} \cdot {[}\!{[}\underline{\boldsymbol\phi}\cdot\mathbf n{]}\!{]} \;\mathrm dS
  \quad \forall\underline{\boldsymbol\phi}
\end{eqnarray}
where $\int_\Omega\underline{\mathbf A}:\underline{\mathbf B}\;\mathrm dx$ denotes the standard
$L^2$ inner product on $\Omega$ for two tensor-valued functions $\underline{\mathbf A}$ and $\underline{\mathbf B}$.
An advantage of the formulation in (\ref{eq:hessian_recovery1})--(\ref{eq:hessian_recovery2})
is that it does not require the control field to have any spatial derivatives at all, meaning it
can be applied to both continuous and discontinuous fields.

Taken together, (\ref{eq:hessian_recovery1}) and (\ref{eq:hessian_recovery2}) give rise to
a $2\times2$ block matrix system.
We solve this system using Schur complement preconditioners and GMRES as the linear solver.
A major advantage of our approach is that the cost function $J$ is defined symbolically, including the computation of the Hessian $\underline{\mathbf H}(\theta)$.
As such, the solution of (\ref{eq:hessian_recovery1})--(\ref{eq:hessian_recovery2}) to approximate the Hessian is automatically taken into account in the computation of the gradient $\mathrm dJ / \mathrm d\bm{\theta}$ during the optimization process.

\section*{Open Research}




The model source code is publicly available. Firedrake, and its components, may be obtained
from \mbox{\url{https://www.firedrakeproject.org}} (last access: 22 June 2023);
Thetis may be obtained from \mbox{\url{http://thetisproject.org}} (last access: 22 June 2023).
The exact software versions used to produce the results in this paper have been archived on Zenodo \cite{zenodo/Firedrake,zenodo/Thetis}.
The model configuration as well as the model and observation time series used to produce the presented results are also available on Zenodo \cite{zenodo/Data}.
Tidal forcing and inversion were computed with the Uptide Python package \cite{zenodo/Uptide}.
Data analysis and visualization were carried out with Matplotlib \cite{hunter2007} and Iris \cite{iris2022} Python packages.

The MEPS atmospheric forecast data can be obtained from the Norwegian Meteorological Institute, \url{https://github.com/metno/NWPdocs/wiki/Data-access} (last access: 22 June 2023).
The ECMWF (European Centre for Medium-Range Weather Forecasts) HRES atmospheric forecast data can be accessed from the ECMWF services, subject to license restrictions, \url{https://www.ecmwf.int/en/forecasts/accessing-forecasts}  (last access: 22 June 2023).
The ERA5 reanalysis atmospheric data can be downloaded from the Copernicus Climate Change Service (C3S) Climate Data Store (CDS) \cite{era5}.
The EHYPE river runoff data can be obtained from the Swedish Meteorological and Hydrological Institute, subject to license restrictions.
The TPXO global tidal atlas (TPXO9-atlas-v1; \citeA{egbert2002}) can be obtained from \url{https://www.tpxo.net} (last access: 22 June 2023).
Bathymetric data is provided by the \citeA{emodnet2020}.
The observational data can be obtained from the Copernicus Marine Service \cite{cmems-baltic,cmems-northsea}.


\acknowledgments
%
%
%
%
%
%


The authors wish to acknowledge CSC – IT Center for Science, Finland, for computational resources.
The work has received support from the Project HPC-EUROPA3 (INFRAIA-2016-1-730897), with the support of the EC Research Innovation Action under the H2020 Programme; in particular, J. G. Wallwork gratefully acknowledges the support of the Finnish Meteorological Institute and the computer resources and technical support provided by CSC.
J. G. Wallwork was funded under the embedded CSE programme of the ARCHER2 UK National Supercomputing Service (\url{http://www.archer2.ac.uk}).
S. C. Kramer acknowledges support from EPSRC under grant EP/R029423/1.
We are grateful to Prof. Matthew Piggott (Imperial College London) for his insightful comments and suggestions.
We thank the four anonymous reviewers for their comments that helped improve the manuscript.


%
%



\bibliography{references}

%
%
%
%
%

\end{document}